\documentstyle[12pt]{article}
\makeatletter
\def\revtex@ver{3.0}
\def\revtex@date{10 Jan 93}
\def\revtex@org{AAS}
\def\revtex@jnl{AAS}
\def\revtex@genre{preprint}
\def\revtex@pageid{\xdef\@thefnmark{\null}
\@footnotetext{This \revtex@genre\space was prepared with the
		   \revtex@org\space \LaTeX\ macros v\revtex@ver.}}
\def\genre@MS{manuscript}
\def\genre@PP{preprint}
\ifx\revtex@genre\genre@PP
\ifnum\@ptsize<1
\fi
\fi
\def\ps@plaintop{\let\@mkboth\@gobbletwo
\def\@oddfoot{}\def\@oddhead{\rm\hfil--\space\thepage\space--\hfil}
\def\@evenfoot{}\let\@evenhead\@oddhead}
\ps@plaintop
\def\@tightleading{1.1}
\def\@doubleleading{1.6}
\def\baselinestretch{\@doubleleading}
\def\tighten{\def\baselinestretch{\@tightleading}}

\def\singlespace{\def\baselinestretch{\@tightleading}\normalsize}
\def\doublespace{\def\baselinestretch{\@doubleleading}\normalsize}
\def\sec@upcase#1{\relax{#1}}
\def\eqsecnum{
\@newctr{equation}[section]
\def\theequation{\hbox{\normalsize\arabic{section}-\arabic{equation}}}}
\tighten
\def\@journalname{The Astropolitical Journal}
\def\cpr@holder{American Astronomical Society}
\def\received#1{\gdef\@recvdate{#1}} \received{\relax}
\def\revised#1{\gdef\@revisedate{#1}} \revised{\relax}
\def\accepted#1{\gdef\@accptdate{#1}} \accepted{\relax}
\def\journalid#1#2{\gdef\@jourvol{#1}\gdef\@jourdate{#2}}
\def\articleid#1#2{\gdef\@startpage{#1}\gdef\@finishpage{#2}}
\def\paperid#1{\gdef\@paperid{#1}} \paperid{MS-0001-SAMP}
\def\ccc#1{\gdef\CCC@code{#1}} \ccc{000-00\$75.95-CDB}
\def\cpright#1#2{\@nameuse{cpr@#1} \gdef\cpr@year{#2}
\typeout{`#1' copyright \cpr@year.}}
\newcount\@cprtype \@cprtype=\@ne
\def\cpr@AAS{\@cprtype=1}
\def\cpr@PD{\@cprtype=2}
\def\cpr@Crown{\@cprtype=3}
\def\cpr@none{\@cprtype=4}
\def\cpr@ASP{\@cprtype=5}
\def\cpr@year{\number\year}
\def\@slug{\par\noindent
\ifcase\@cprtype
	\relax
\or
	Copyright \cpr@year\space by the \cpr@holder.
\or
	This article is in the public domain.
\or
	Crown copyright \cpr@year\space by the \cpr@holder.
\or
	No copyright is claimed for this article.
\or
	Copyright \cpr@year\space by the ASP.
\fi
\par\noindent
Manuscript number \@paperid.\par\noindent
\CCC@code
}
\def\lefthead#1{\gdef\@versohead{#1}} \lefthead{\relax}
\def\righthead#1{\gdef\@rectohead{#1}} \righthead{\relax}
\def\@runheads{\@tempcnta\c@page
\@whilenum \@tempcnta >0\do{
\vskip 3ex
\hbox to30pc{\small\expandafter\uppercase\expandafter{\@versohead}:
	\expandafter\uppercase\expandafter{\@rectohead}\hfil}
\advance\@tempcnta by\m@ne}
}
\def\slugcomment#1{\gdef\slug@comment{#1}} \slugcomment{}
\newdimen\@slugcmmntwidth \@slugcmmntwidth .67\textwidth
\long\def\@makeslugcmmnt{\ifx\slug@comment\@empty\relax\else
\setbox\@tempboxa\hbox{\slug@comment}
\ifdim \wd\@tempboxa >\@slugcmmntwidth
\hbox to\textwidth{\hss
	    \parbox\@slugcmmntwidth\slug@comment}
\else
\hbox to\textwidth{\hfil\box\@tempboxa}
\fi
\vskip 2ex
\fi}
\def\@rcvaccrule{\vrule\@width1.75in\@height0.5pt\@depth\z@}
\def\@dates{{Received}\space%
\if\@recvdate\relax\@rcvaccrule\else\@recvdate\fi;%
\hspace{1.5em}{accepted}\space%
\if\@accptdate\relax\@rcvaccrule\else\@accptdate\fi%
}
\def\sluginfo{{\center
\@dates

\endcenter}}

\def\abstract{
\begin{center}
{\bf{ABSTRACT}}
\end{center}
\quotation
}
\def\title#1{\@makeslugcmmnt{\center\large\bf{#1}\endcenter}
\thispagestyle{empty}}
\def\author#1{{\topsep\z@\center\normalsize#1\endcenter}}
\let\authoraddr=\@gobble
\def\affil#1{\vspace*{-2.5ex}{\topsep\z@\center#1\endcenter}}

\def\and{\vspace*{-0.5ex}{\topsep\z@\center and\endcenter}}
\def\@keywordtext{Subject headings}
\def\@keyworddelim{---}
\def\keywords#1{\vspace*{-.7ex}
\if@twocolumn\noindent{{\it\@keywordtext:\/}\space\@kwds{#1}}
\else{\quote{\it\@keywordtext:\/}\space\@kwds{#1}\endquote}
\fi}

\def\@kwds#1{#1\relax}
\skip\footins 4ex plus 1ex minus .5ex
\newif\if@firstsection \@firstsectiontrue
\def\section{\if@firstsection
\@firstsectionfalse\fi
\@startsection {section}{1}{\z@}
{5ex plus .5ex}{1ex plus .2ex}{\normalsize\bf}}
\def\subsection{\@startsection{subsection}{2}{\z@}
{5ex plus .5ex}{1ex plus .2ex}{\normalsize\bf}}
\def\subsubsection{\@startsection{subsubsection}{3}{\z@}
{5ex plus .5ex}{1ex plus .2ex}{\normalsize\it}}
\def\thesection{\@arabic{\c@section}.}
\def\thesubsection{\thesection\@arabic{\c@subsection}.}
\def\thesubsubsection{\thesubsection\@arabic{\c@subsubsection}.}
\def\theparagraph{\thesubsubsection\@arabic{\c@paragraph}:}

\def\@sect#1#2#3#4#5#6[#7]#8{\ifnum #2>\c@secnumdepth
\def\@svsec{}\else
\refstepcounter{#1}\edef\@svsec{\csname the#1\endcsname\hskip 1em }\fi
\@tempskipa #5\relax
\ifdim \@tempskipa>\z@
\begingroup \center#6\relax
\@hangfrom{\hskip #3\relax\@svsec}{\interlinepenalty \@M
	  \sec@upcase{#8}\par}%
\endcenter\endgroup
\csname #1mark\endcsname{#7}\addcontentsline
{toc}{#1}{\ifnum #2>\c@secnumdepth \else
\protect\numberline{\csname the#1\endcsname}\fi
#7}\else
\def\@svsechd{#6\hskip #3\@svsec \sec@upcase{#8}\csname #1mark\endcsname
{#7}\addcontentsline
{toc}{#1}{\ifnum #2>\c@secnumdepth \else
\protect\numberline{\csname the#1\endcsname}\fi
#7}}\fi
\@xsect{#5}}
\def\@ssect#1#2#3#4#5{\@tempskipa #3\relax
\ifdim \@tempskipa>\z@
\begingroup #4\center\@hangfrom{\hskip #1}{\interlinepenalty \@M
\sec@upcase{#5}\par}\endcenter\endgroup
\else \def\@svsechd{#4\hskip #1\relax \sec@upcase{#5}}\fi
\@xsect{#3}}
\def\appendix{\par
\setcounter{section}{0}
\setcounter{subsection}{0}
\setcounter{equation}{0}
\def\thesection{\Alph{section}.}
\def\theequation{\hbox{\normalsize\Alph{section}\arabic{equation}}}}
\newcounter{cureqno}
{\let\theequation\@curtheeqn%
\setcounter{equation}{\value{cureqno}}}
\def\eqnum#1{\def\theequation{#1}\let\@currentlabel\theequation
\addtocounter{equation}{\m@ne}}
\def\references{\subsection*{REFERENCES}
\bgroup\parindent=\z@\parskip=\itemsep
\def\refpar{\par\hangindent=3em\hangafter=1}}
\def\endreferences{\refpar\egroup\revtex@pageid}
\def\thebibliography{\subsection*{REFERENCES}
\list{\null}{\leftmargin 3em\labelwidth\z@\labelsep\z@\itemindent -3em
\usecounter{enumi}}
\def\refpar{\relax}
\def\newblock{\hskip .11em plus .33em minus .07em}
\sloppy\clubpenalty4000\widowpenalty4000
\sfcode`\.=1000\relax}
\def\endthebibliography{\endlist\revtex@pageid}
\def\@biblabel#1{\relax}
\def\@cite#1#2{#1\if@tempswa , #2\fi}

\def\@citex[#1]#2{\if@filesw\immediate\write\@auxout{\string\citation{#2}}\fi
\def\@citea{}\@cite{\@for\@citeb:=#2\do
{\@citea\def\@citea{,\penalty\@m\ }\@ifundefined
{b@\@citeb}{\@warning
{Citation `\@citeb' on page \thepage \space undefined}}%
{\csname b@\@citeb\endcsname}}}{#1}}

\newtoks\@temptokenb
\def\tblnote@list{}
\def\tablenotetext#1#2{
\@temptokena={\vspace{.5ex}{\noindent\llap{$^{#1}$}#2}\par}
\@temptokenb=\expandafter{\tblnote@list}
\xdef\tblnote@list{\the\@temptokenb\the\@temptokena}}
\def\spew@tblnotes{
\ifx\tblnote@list\@empty\relax
\else
\vspace{4.5ex}
\footnoterule
\vspace{.5ex}
{\footnotesize\tblnote@list}
\gdef\tblnote@list{}
\fi}
\def\endtable{\spew@tblnotes\end@float}
\@namedef{endtable*}{\spew@tblnotes\end@dblfloat}

\long\def\@makecaption#1#2{\vskip 2ex\noindent #1 #2\par}
\def\tablenum#1{\def\thetable{#1}\let\@currentlabel\thetable
\addtocounter{table}{\m@ne}}
\def\figurenum#1{\def\thefigure{#1}\let\@currentlabel\thefigure
\addtocounter{figure}{\m@ne}}
\newbox\pt@box
\newdimen\pt@width
\newcount\pt@line
\newcount\pt@nlines
\newcount\pt@ncol
\def\colhead#1{\omit\hidewidth{#1}\hidewidth\global\advance\pt@ncol by\@ne}
\def\tablecaption#1{\gdef\pt@caption{#1}} \def\pt@caption{\relax}
\def\tablehead#1{\gdef\pt@head{\hline\hline\relax\\[-1.7ex]
#1\hskip\tabcolsep\\[.7ex]\hline\relax\\[-1.5ex]}} \def\pt@head{\relax}
\def\tabletail#1{\gdef\pt@tail{#1}} \def\pt@tail{\relax}
\def\tablewidth#1{\pt@width=#1} \pt@width\textwidth
\def\tableheadfrac#1{\gdef\pt@headfrac{#1}} \def\pt@headfrac{.1}
\def\pt@calcnlines{\@tempdima\pt@headfrac\textheight
\@tempdimb\textheight\advance\@tempdimb by-\@tempdima
\@tempdima\arraystretch\baselineskip
\divide\@tempdimb by\@tempdima
\global\pt@nlines\@tempdimb}
\def\pt@tabular{\hbox \bgroup $\let\@acol\@ptabacol
\let\@classz\@tabclassz
\let\@classiv\@tabclassiv \let\\\@tabularcr\@tabarray}
\def\@ptabacol{\edef\@preamble{\@preamble \hskip \tabcolsep\tabskip\fill}}
\def\fnum@ptable{Table \thetable}
\def\fnum@ptablecont{Table \thetable---{\rm Continued}}
\def\set@mkcaption{\long\def\@makecaption##1##2{
\center\rm##1.\quad##2\endcenter\vskip 2.5ex}}
\def\set@mkcaptioncont{\long\def\@makecaption##1##2{
\center\rm##1\endcenter\vskip 2.5ex}}
{\crcr\noalign{\vskip .7ex}\hline\endtabular%
\pt@width\wd\pt@box\box\pt@box\spew@ptblnotes%
\typeout{Table \thetable\space has been set to width \the\pt@width}%
\endcenter\end@float}
\def\startdata{\pt@line=0\pt@calcnlines%
\ifdim\pt@width>\z@\def\@halignto{to \pt@width}\else\def\@halignto{}\fi%
\let\fnum@table=\fnum@ptable\set@mkcaption%
\@float{table}\center\caption{\pt@caption}\leavevmode%
\setbox\pt@box=\pt@tabular{\pt@format}\pt@head}
\def\pt@nl{\global\advance\pt@line by\@ne%
\ifnum\pt@line=\pt@nlines%
\endtabular\box\pt@box
\endcenter\end@float\clearpage%
\addtocounter{table}{\m@ne}%
\let\fnum@table=\fnum@ptablecont\set@mkcaptioncont%
\@float{table}\center\caption{\pt@caption}\leavevmode%
\global\pt@ncol=0%
\setbox\pt@box=\pt@tabular{\pt@format}\pt@head%
\global\pt@line=0%
\else\\
\fi}
\let\nl=\pt@nl
\let\nextline=\pt@nl

\def\tablebreak{\pt@line\pt@nlines\advance\pt@line by\m@ne\pt@nl}
\def\cutinhead#1{\noalign{\vskip 1.5ex}
\hline\pt@nl\noalign{\vskip -2.0ex}
\multicolumn{\pt@ncol}{c}{#1}\pt@nl
\noalign{\vskip .8ex}
\hline\pt@nl\noalign{\vskip -2ex}}
\def\sidehead#1{\noalign{\vskip 1.5ex}
\multicolumn{\pt@ncol}{@{\hskip\z@}l}{#1}\pt@nl
\noalign{\vskip .5ex}}
\def\set@tblnotetext{\def\tablenotetext##1##2{{%
\@temptokena={\vspace{0ex}{%
\parbox{\pt@width}{\hskip1em$^{\rm ##1}$##2}\par}}%
\@temptokenb=\expandafter{\tblnote@list}
\xdef\tblnote@list{\the\@temptokenb\the\@temptokena}}}}
\def\spew@ptblnotes{
\ifx\tblnote@list\@empty\relax
\else
\par
\vspace{2ex}
{\tblnote@list}
\gdef\tblnote@list{}
\fi}
\def\tablerefs#1{\@temptokena={\vspace*{3ex}{%
\parbox{\pt@width}{\hskip1em\rm References. --- #1}\par}}%
\@temptokenb=\expandafter{\tblnote@list}
\xdef\tblnote@list{\the\@temptokenb\the\@temptokena}}
\def\tablecomments#1{\@temptokena={\vspace*{3ex}{%
\parbox{\pt@width}{\hskip1em\rm Note. --- #1}\par}}%
\@temptokenb=\expandafter{\tblnote@list}
\xdef\tblnote@list{\the\@temptokenb\the\@temptokena}}
\def\thefigure{\@arabic\c@figure}
\def\fnum@figure{{\rm Fig.\space\thefigure.---}}
\def\thetable{\@arabic\c@table}
\def\fnum@table{{\rm Table \thetable:}}
\def\fps@figure{bp}
\def\fps@table{bp}
\@ifundefined{epsfbox}{\@input{epsf.sty}}{\relax}

\let\jnl@style=\rm
\def\ref@jnl#1{{\jnl@style#1}}
\def\aj{\ref@jnl{AJ}}
\def\araa{\ref@jnl{ARA\&A}}
\def\apj{\ref@jnl{ApJ}}
\def\apjl{\ref@jnl{ApJ}}
\def\apjs{\ref@jnl{ApJS}}
\def\ao{\ref@jnl{Appl.Optics}}
\def\apss{\ref@jnl{Ap\&SS}}
\def\aap{\ref@jnl{A\&A}}
\def\aapr{\ref@jnl{A\&A~Rev.}}
\def\aaps{\ref@jnl{A\&AS}}
\def\azh{\ref@jnl{AZh}}
\def\baas{\ref@jnl{BAAS}}
\def\jrasc{\ref@jnl{JRASC}}
\def\memras{\ref@jnl{MmRAS}}
\def\mnras{\ref@jnl{MNRAS}}
\def\pra{\ref@jnl{Phys.Rev.A}}
\def\prb{\ref@jnl{Phys.Rev.B}}
\def\prc{\ref@jnl{Phys.Rev.C}}
\def\prd{\ref@jnl{Phys.Rev.D}}
\def\prl{\ref@jnl{Phys.Rev.Lett}}
\def\pasp{\ref@jnl{PASP}}
\def\pasj{\ref@jnl{PASJ}}
\def\qjras{\ref@jnl{QJRAS}}
\def\skytel{\ref@jnl{S\&T}}
\def\solphys{\ref@jnl{Solar~Phys.}}
\def\sovast{\ref@jnl{Soviet~Ast.}}
\def\ssr{\ref@jnl{Space~Sci.Rev.}}
\def\zap{\ref@jnl{ZAp}}

\let\apjlett=\apjl

\def\lesssim{\mathrel{\hbox{\rlap{\hbox{\lower4pt\hbox{$\sim$}}}\hbox{$<$}}}}
\def\gtrsim{\mathrel{\hbox{\rlap{\hbox{\lower4pt\hbox{$\sim$}}}\hbox{$>$}}}}

\def\ion#1#2{#1$\;${\small\rm\@Roman{#2}}\relax}

\newcount\lecurrentfam
\def\LaTeX{\lecurrentfam=\the\fam \leavevmode L\raise.42ex
\hbox{$\fam\lecurrentfam\scriptstyle\kern-.3em A$}\kern-.15em\TeX}
\def\sizrpt{
(\fontname\the\font): em=\the\fontdimen6\font, ex=\the\fontdimen5\font
\typeout{
(\fontname\the\font): em=\the\fontdimen6\font, ex=\the\fontdimen5\font
}}
\makeatother
 at 14.40truept
 at 17.28truept

\def \einstein{\hbox {\it Einstein \/}}
\def\heading#1{\goodbreak\medskip\centerline{\bf #1}\medskip}
\def\section#1{\goodbreak\medskip\centerline{\bf #1}\medskip}
\def \bref{\par \noindent \hangindent=1.5 truecm \hangafter=1} 	% References
\newcount\itemno
\def \ino         { \the\itemno\global\advance\itemno by 1 }
\def \ii          { \itemitem }

\def \mpc	{{\rm\ Mpc}}
\def \kpc	{{\rm\ kpc}}

\def \pspc	{{\rm\ PSPC}}
\def \kms       {\hbox{ km s$^{-1}$}}

\def \ie	{\hbox{\it i.e.}}
\def \eg	{\hbox{\it e.g.}}
\def \cf	{\hbox{\it cf.}}

\def \etal 	{{\it et al.\ }}

\def \K		{ \hbox{$\,$ K} }
\def \Ho	{{\rm\ H_{o}}}
\def \kmsmpc	{{\rm\ km\ s^{-1}\ Mpc^{-1}}}
\def \kev	{{\rm\ keV}}
\def \msol	{{\rm M}_\odot}
\def \lsol	{{\rm L}_\odot}

\def \h50inv	{\hbox{$\, h^{-1}_{50}$} }
\def \ergs	{ \hbox{$\,$ erg s$^{-1}$} }
\def \ctss	{ \hbox{$\,$ cts s$^{-1}$} }

\def \cgsflux	{{\rm\ erg\ s^{-1}\ cm^{-2}}}
\def \sbunits	{{\rm\ cts\ s^{-1}\ arcmin^{-2}}}
\def \cc	{{\rm\ cm^{-3}}}
\def \se	{\!=\!}
\def \ssim	{\! \sim \!}

\def \sequiv	{\! \equiv \!}
\def \spropto	{\! \propto \!}
\def\aj{{\it A. J.\/}}
\def\annrev{{\it Ann. Rev. Astr. Astrophys.\/}}
\def\apj{{\it Ap. J.\/}}
\def\apjlett{{\it Ap. J. Lett.\/}}
\def\apjsup{{\it Ap. J. Suppl.\/}}
\def\aa{{\it Astron. Astrophys.\/}}
\def\baas{{\it Bull. Am. Astr. Soc.\/}}

\def\mnras{{\it M.N.R.A.S.\/}}
\def\nature{{\it Nature \/}}
\def\pasp{{\it Publ. Astron. Soc. Pac.\/}}
\def\prl{{\it Phys. Rev. Lett.\/}}
\def\prd{{\it Phys Rev. D\/}}
\def\progphys{{\it Prog. Theor. Phys.\/}}
\def\rmp{{\it Rev. Mod. Phys.\/}}
\def\\{\hfil\break}
\def\spose#1{\hbox to 0pt{#1\hss}}
\def\lta{\mathrel{\spose{\lower 3pt\hbox{$\mathchar"218$}}
     \raise 2.0pt\hbox{$\mathchar"13C$}}}
\def\gta{\mathrel{\spose{\lower 3pt\hbox{$\mathchar"218$}}
     \raise 2.0pt\hbox{$\mathchar"13E$}}}

\def\beq{\begin{equation}}
\def\eeq{\end{equation}}
\begin{document}
\begin{titlepage}

% \begin{center}

\title {A SIMULATION OF THE INTRACLUSTER MEDIUM \\
 WITH FEEDBACK FROM CLUSTER GALAXIES}

\author {Christopher A. Metzler and August E. Evrard}
\medskip
\affil {Department of Physics, University of Michigan, Ann Arbor, MI 48109}
\medskip
\centerline{(to be submitted to the {\it Astrophysical Journal})}

% \end{center}

\begin{abstract}

We detail method and report first results from a three--dimensional
hydrodynamical and N--body simulation of the formation and evolution of
a Coma--sized cluster of galaxies, with the intent of studying
the history of the hot, X--ray emitting intracluster medium.
Cluster gas, galaxies, and
dark matter are included in the model.  The galaxies and dark matter
feel gravitational forces; the cluster gas also undergoes hydrodynamical
effects such as shock heating and PdV work.  For the first time in
three dimensions, we include modelling of ejection of processed gas
from the simulated galaxies by winds,
including heating and heavy element enrichment.  For comparison, we
employ a `pure infall' simulation using the same initial conditions
but with no galaxies or winds.

The galactic feedback raises the entropy of the intracluster gas,
preventing it from collapsing to densities as high as those attained
in the infall model.  This effect is more pronounced in
sub--clusters formed at high redshift.  The cluster with feedback
is always less X--ray luminous but experiences more rapid luminosity
evolution than the pure infall cluster.

The final gas temperature
in the model with feedback is $\ssim 15\%$ larger than in the infall
model and is very near to isothermal within $1.5 \mpc$.
The cluster galaxies in the feedback model have a velocity
dispersion $\ssim 15\%$ lower than the dark matter.  This results in
the true ratio of specific energies in galaxies to gas being
less than one, $\beta_{spec} \ssim 0.7$.  The infall model predicts
$\beta_{spec} \ssim 1.2$.

The morphology of the X--ray emission is little affected by
feedback.  The emission profiles of both clusters are well described
by the standard $\beta$--model, with $\beta_{fit} \simeq 0.7-0.9$.
X--ray mass estimates based on the assumptions of hydrostatic
equilibrium and the applicability of the $\beta$--model are quite
accurate in both cases.

A strong iron abundance gradient is present, which develops as a
consequence of the steepening of the galaxy density profile over time.
Spectroscopic observations using non--imaging detectors with wide
($\ssim 45^\prime$) fields of view dramatically smear the gradient.
Observations with arcminute resolution, made available with the
ASCA satellite, would readily resolve the gradient.  We predict such
observations should also resolve inhomogensous structure in the cluster
iron abundance.  However, the amplitudes of both the radial gradient and
the small--scale inhomogeneities are sensitive to uncertain details
of wind models from early--type galaxies.
\end{abstract}

\bigskip\keywords{Galaxies-clusters, cosmology-theory}

\end{titlepage}

\section{I. Introduction}

The space between galaxies in clusters is filled with a
hot, X-ray emitting plasma known as the {\it intracluster medium},
or ICM.  The study of the origin and evolution of this gas is
guided by three observations.  The ratio of the mass in the ICM to
the mass in cluster galaxies, $M_{ICM} / M_{stars}$, seems to increase
from $\sim 1$ for poorer clusters to $\sim 4-6$ for rich clusters
(\cf\/ David \etal 1990a); that
this ratio is greater than one implies that the majority of the gas
has not been processed through cluster galaxies.  On the other hand,
spectroscopic observations of clusters show iron abundances in the ICM
of $\sim 0.5$ solar, implying that at least some of the ICM has been
cycled through stars.  Finally, the ratio of specific energy in cluster
galaxies to specific energy in cluster gas, the quantity
$\beta$ discussed below, seems to be less than one for a
majority of systems (\cf\/ Edge \& Stewart 1991), implying more energy per
unit mass in gas than in galaxies.  Any good theory of the origin and
evolution of the ICM must account for these observations.

The paradigm for the formation of clusters and the ICM involves
the infall of galaxies and gas into a hierarchically developing potential
well (Gunn \& Gott 1972).  The combination of gravitational collapse and
shock heating drives the gas up to observed temperatures
($T \sim 10^{7}-10^{8} K$).  Such a scenario can account for the
huge amount of gas in clusters, but cannot immediately account for
the presence of metals in the gas.  There are three sources for the
processed gas in the ICM commonly considered:  {\it i)} a `background'
metallicity in the infalling gas, perhaps from Population III stars
(Carr, Bond, \& Arnett 1984); {\it ii)} processed gas ram pressure
stripped out of cluster galaxies by the existing ICM (Gunn \& Gott 1972;
Biermann 1978; Takeda, Nulsen, \& Fabian 1984; Gaetz, Salpeter,
\& Shaviv 1987); or {\it iii)}
processed gas driven out of cluster galaxies by energetic winds.

If cluster metallicities originate predominantly from
abundances in the gas that existed before the cluster formed, then,
neglecting settling of heavy ions, one would expect no radial gradient
in metallicity in the cluster.  One would also expect cluster metal
abundances to be independent of the mass of the cluster.  While observed
iron abundances in rich clusters lie within a fairly narrow range, there
does exist some evidence for a decrease in metal abundance with
cluster mass (\cf\/ Jones \& Forman 1992).  If the metals come from
processed gas that has been stripped out of cluster galaxies, then
one would expect a strong peak in metallicity near the center, where
the density of the ICM is the largest.

Ejection by galactic winds was originally considered as
a source for the ICM (Yahil \& Ostriker 1973; Larson and
Dinerstein 1975).  As a galaxy evolves, supernova activity affects
the state of the interstellar medium.  Subsonic galactic
outflows or transonic galactic winds can develop, which in turn eject hot,
processed gas out of the galaxy. Winds have been observed in starburst
galaxies (\cf\/ McCarthy, Heckman, \& van Bruegel 1987; Heckman, Armus,
\& Miley 1987, 1990), and winds and
outflows are general properties of models of the evolution of ellipticals
(Matteucci \& Tornamb\`{e} 1987; Matteucci \& Vettolani 1988;
David, Forman, \& Jones 1990b; Ciotti \etal 1991; David, Forman,
\& Jones 1991b).
It seems reasonable to assume that {\it some} feedback of processed gas
from cluster galaxies into the ICM would take place.  A distinguishing
characteristic of the wind scenario is that it results in not only metals,
but also energy, being injected into the ICM.  The typical specific energies
associated with supernova--driven galactic winds are on the order of, or
greater than, the specific energies associated with infall into a rich
cluster, possibly resulting in greater specific energy for cluster gas than
one finds for cluster galaxies.  Winds thus provide a simple mechanism
for explaining $\beta \,<\, 1$.

Simulations are necessary to accurately model cluster evolution
and test these hypotheses.  In the 1970's, various papers studied the
formation and evolution of an X--ray cluster (\eg\/ Gull \& Northover
1975; Lea 1976, Takahara \etal 1976; Cowie \& Perrenod 1978; Perrenod
1978).  The simulations associated with these papers were one--dimensional
hydrodynamical models, usually with fixed, static gravitational
potentials.  Perrenod (1978) introduced a time--varying potential
into his one-dimensional simulations which was extracted from the
N--body simulation of White (1976).  Cowie \& Perrenod (1978) and
Perrenod (1978) also considered models in which the entire ICM originated
from within cluster galaxies, assuming some static, or simply varying in
space and time, mass ejection rate $\phi(r,t)$.  Evrard (1990a,b)
performed three--dimensional simulations of X--ray clusters where, rather
than adopting some {\it ad hoc} gravitational potential, both collisionless
dark matter and collisional baryonic fluids were included; the baryons
fell into the developing dark matter potential well and underwent
shock heating.  Thomas \& Couchman (1992), Katz \& White (1993),
Tsai, Katz, \& Bertschinger (1993) and Navarro, Frenk \& White (1993)
 also studied similar
three--dimensional models with somewhat higher numerical resolution.
This paper extends the work of Evrard (1990b), improving resolution and
allowing examination of the effects of galactic winds upon the intracluster
medium.  In particular, we wish to compare the evolution of a simulated
cluster with and without energetic ejection from galaxies, to isolate
any changes originating from the additional physics of galactic feedback.
Future papers will contrast winds with gas--dynamical stripping, and
will allow use of an ensemble of simulated clusters to explore
correlations of observables as well as X--ray and optical
indicators of dynamical state.

This paper is organized as follows.  Section II describes the
numerical method used in modelling feedback from galaxies.  The specific
cluster model studied for this paper, and the results of the simulations,
are described in Section III.  Section IV considers what these simulations
can tell us about cluster mass estimates, the cluster baryon fraction
and the specific energies of various cluster mass components.
Finally, Section V summarizes the results, lists our conclusions,
and presents ideas for further study.

\section{II.  Method}

\heading{a) Cosmological Model, Initial Conditions, and Simulation Algorithm}

The parameters for our models follow those of the standard
biased cold dark matter (CDM) scenario (Blumenthal \etal 1984; Davis
\etal 1985):  $\Omega = 1$, baryonic fraction $\Omega_{b} = 0.1$,
Hubble constant $h \se 0.5$ (where $h \sequiv \Ho/100 \kmsmpc$)
and the spectrum is normalized to
have a linear evolved variance in $16 \mpc$ spheres of
$\sigma_{8}\,=\,0.59$,  equivalent to a bias factor $b \sequiv
\sigma_8^{-1} \se 1.7$.
An initial density field is generated in a $40 \mpc$ cube from a
CDM power spectrum
using the path integral method of Bertschinger (1987), which
constrains the field smoothed with a Gaussian filter to have a
specified value at the center of the simulated volume.
For this model, we smooth over a length scale of
$R_{f}\,=\,8 \mpc$, corresponding to a mass scale of
$M_{f}\,=\,\left(2\pi\right)^{3/2}R_{f}^3\;\rho \,=\,
5.6 \times 10^{14} \msol$ (Bardeen \etal 1986).
The perturbation height at the center was constrained to
a value $\delta_{c} \,=\, 2.0$ when filtered on scale $M_{f}$, resulting
in a $3.6 \sigma$ peak on this scale near the center of the box.
The primordial density field is used to generate a particle
distribution  at the starting redshift using the Zel'dovich
approximation, as described in Efstathiou \etal (1985).

Our simulation algorithm P3MSPH allows us to track the evolution of
both collisionless dark matter and collisional baryonic fluids.  The
particles of all fluids are coupled together by gravity; the baryonic
fluid also feels gas forces.  The effects of gravity are modelled by the
well known particle-particle--particle-mesh ($P^{3}M$) algorithm of
Efstathiou \& Eastwood (1981); a detailed explanation is in Hockney \&
Eastwood (1981). The hydrodynamical terms in the force equations are solved
using the Smoothed Particle Hydrodynamics (SPH) algorithm of Gingold and
Monaghan (1977).  A review of SPH can be found in Monaghan (1992).
The combined algorithm is described in Evrard (1988) and a description
of some of the analysis techniques used here is provided in Evrard (1990b).

\heading{b) Inclusion of Galaxies and Ejection of Gas}

Previous efforts to model the effects on the ICM of ejection of
gas from galaxies were limited by their assumed spherical symmetry (\eg\/
Cowie \& Perrenod 1978; Perrenod 1978).  Mass
and energy ejection was typically parametrized by some {\it ad hoc} mass
injection rate $\phi (r,t)$, $r$ being the radial distance
from the center of the cluster.  The three--dimensional capability of our
simulation code allows us to consider time--dependent mass ejection rates
which truly vary with position.  Since cluster galaxies are the source of
this processed gas, the time--evolution of the 3-D mass ejection rate
$\phi (\vec{r},t)$ depends on both the evolution of the spatial distribution
of galaxies and the evolution of the galaxies themselves.  Therefore,
in addition to simulating background fluids of collisionless dark matter
and collisional baryonic gas, we also simulate a third set of particles
representing individual cluster galaxies.  The galaxy phase--space
distribution evolves consistently with the other two fluids, being coupled
to them by gravity.  This method offers another benefit; by modelling the
galaxy motions explicitly, we can study the dynamics of the ensemble of
galaxies as well as the evolution of the ICM.
Analyses of cluster galaxy kinematics can be done without having to make the
assumption that the evolution of the galaxy distribution is
well--approximated by the evolution of the dark matter distribution. In
particular, issues of velocity bias, anisotropies in galaxy velocity
dispersions, and optical substructure can be addressed.

Ideally, our simulation would include the formation of
individual galaxies, which would then be the source of ejected
material.  However, simulations which can model galaxy formation
in such a cosmological context require a large amount of computational
overhead.  Less numerically ambitious simulations will allow us to generate
many runs and perform statistical analyses, but we consequently cannot
adequately resolve scales relevant for galaxy formation.  We therefore
approximate what we cannot resolve by putting galaxies in `by hand',
at locations corresponding to peaks in the initial density field.
The correspondence between peaks and subsequent collapsed objects has
been investigated in N-body experiments by Frenk \etal (1988) and
Katz, Quinn \& Gelb (1992).  These studies have shown that high
peaks are very likely to pick out sites which later collapse into
bound structures, but the correspondence is very poor for low peaks.
We thus limit our analysis to high peaks in an attempt to approximate
the behavior of  L$^{\ast}$ galaxies.  It should be kept in
mind that this is an approximate treatment, intended to generate
the observed number density of bright galaxies in a manner consistent
with their statistical clustering properties (Davis \etal 1985).
While the method is certainly artificial, it is not
entirely without basis.  Evrard, Silk, \& Szalay (1990), for instance,
found that an identification of peaks with galaxies, combined with a simple
mapping between peak heights and galaxy morphology, could reproduce the
observed morphology--density relation (\eg\/ Dressler 1980).

To locate sites for galaxy insertion, we filter the initial density field
on a scale appropriate for massive, L$^{\ast}$ galaxies.  We use
Gaussian filtering with $R_{f}\,=\,0.5 \mpc$ and identify sites
above a threshold of $2.5\sigma$ as locations for massive galaxy
formation (Davis \etal 1985).  We then return to the
initial particle distribution and replace the gas particles
associated with each peak with a composite ``galaxy particle.''
% The mass of each
% galaxy particle is equal to the number of gas particles removed at that
% peak.
The initial linear momentum of a galaxy particle is set by demanding
conservation of linear momentum when the gas particles are removed.
Applying this technique yielded  108 ``L$^{\ast}$ galaxies'', with an
average mass of $5.06 \times 10^{11} \msol$.  The number density, for
a comoving 40$ \mpc$ box, is then $1.69 \times 10^{-3} \, \mpc^{-3}$,
very similar to  the most recent estimate of the number density of
bright galaxies in the Stromlo--APM survey
$\phi^{\ast}\,=\,1.75\times10^{-3} \, \mpc^{-3}$ (Loveday \etal 1992).

The modelling of galactic winds requires several
approximations.   The formation, duration, intensity, and composition
of such winds are highly dependent upon the assumed stellar initial mass
function (IMF) and star formation rate (SFR), as these determine stellar
mass loss rates, as well as rates of Type I and Type II supernovae.  For
instance, David \etal (1990b, 1991a) and Ciotti \etal (1991) used
hydrodynamical simulations to explore a variety of models of
elliptical galaxies with different IMFs and SFRs; the wind
solutions obtained had similar general properties, but varied strongly
in details.  There is no `most general case' appropriate for our
use.  We have thus coded a general method capable of handling a
variety of possible wind models.  To parametrize the behavior of
the individual galaxies, the simulation code takes time--dependent
specific mass, energy, and iron mass ejection rates as input.  Such
rate curves in reality depend in a complicated fashion on the state
of the interstellar medium.

For each galaxy, as a model
advances in timesteps, the specific mass ejection rate curve is integrated
in time until the ejected mass is equal to the mass of a gas particle.  At
that point, a new gas particle is created at the location of the galaxy
particle, and the mass of the galaxy particle is decremented by one.  The
energy and iron mass fraction of the new gas particle are determined by
integrating those curves over the same time period.

There are three naive concerns with this algorithm.  It
is unclear from galaxy evolutionary models what fraction of the
energy of the wind should be in mechanical energy and what should be
in thermal energy.  Second, an actual galactic outflow ejects a continuous
stream of matter which interacts with the large surrounding volume of gas.
In contrast, the discrete particle and discrete timestep nature of our
simulations necessitates that lumps of gas (individual gas particles)
be ejected periodically, with a large total energy.  We must ensure
that the global properties of the ejected matter, and its effects upon
the rest of the ICM, are not strongly sensitive to the discreteness
limits of our ejection procedure.  Finally, the SPH algorithm is inaccurate
and unstable when hydrodynamic properties vary strongly on the scale of one
particle, as could happen in a poorly--modelled ejection process.

All three of these issues are resolved by addressing the
second.  Modelling a continuous flow as a discrete particle neglects
the interaction with the surrounding medium that would have taken place
in the time up to when the particle is ejected.  We therefore adopt
a paradigm for what happens to ejected gas, and force such interaction
to take place at the time of particle ejection.  We mimic the mixing
which takes place on unresolved scales by smoothing the ejected energy
and momentum with the $N$ nearest particles of the ejected particle.
We use $N \se 30$, though tests showed that the
characteristics of runs on resolved scales were insensitive
to varying $N$ between 10 and 30.
During the runs, we continuously compared the distance over
which we forced interaction
with the sound speed in the ejected gas, to verify that the interactions
we have inserted by--hand do not violate causality bounds.

The paradigm for how the wind--ICM interaction is forced
is as follows.  In the rest frame of a galaxy, we assume its outflow
to be isotropic; then no momentum is associated with that outflow, and
we can define a net momentum of our interacting system to be just the
sum of the momenta of the $N$ neighbors:
\beq
\vec{p}_{tot} = \sum_{i=1}^{N} \vec{p}_{i}.
\eeq
Upon meeting the surrounding medium, the ejected gas particle should gain
momentum from its neighbors; but the total momentum of the ejected
particle plus neighbors, $\vec{p}_{tot}$, should be conserved.
Then, in the rest frame of the ejecting galaxy, the ejected
gas particle's initial momentum is chosen to be the average of the
resulting flow:
\beq
\vec{p}_{ej} = \frac{\vec{p}_{tot}}{N+1}.
\eeq
To bring the ejected particle up to this momentum (and to conserve
net momentum), each neighbor is assumed to have lost $1/N$ the momentum
given to the ejected particle:
\beq
\vec{p}_{i}\,^{\prime} = \vec{p}_{i} - \frac{\vec{p}_{ej}}{N}.
\eeq
Such a process conserves momentum, but it does not conserve energy.
A brief calculation shows that an amount of kinetic energy is lost
equal to
\beq
E_{lost} = \frac{\parallel\vec{p}_{tot}\parallel^{2}}{2mN(N+1)}
\eeq
with $m$ the mass of a gas particle.  But this kinetic energy loss
is not a problem.  In the galaxy's frame, as the ejected gas encounters
the local flow of the intracluster medium and momentum is transferred,
one expects kinetic energy to be converted into heat; this is an
inelastic process.  The energy $E_{lost}$ above is thus added to the
thermal energy we obtained by integrating the wind luminosity curve;
this is the total thermal energy available from the ejection process.
To complete the consideration of the interaction of the wind with its
surroundings, the thermal energies of the ejected particle and the $N$
neighbors are adjusted to the same, average value of the total thermal
energy available divided by $N+1$; in effect, the ejected particle is
created in thermal equilibrium with its surroundings.

Globally, energy is not conserved, as the
energy associated with the new gas particle supposedly came from nuclear
processes within the galaxy, and such sources of energy are not considered
in summing the total kinetic, potential, and thermal energy in the
simulation volume.  However, except for the ``new energy'' associated
with the wind luminosity, all existing energy and momentum is conserved
in this interaction process.

It should be noted that there are relevant physical effects
which are not included in this work.  Our code is equipped to include
the effects of radiative cooling; but we have ignored cooling because
of our poor resolution on scales where cooling should be important.
We expect to include cooling in future runs with higher resolution.  We
also do not include gas--dynamical stripping from cluster galaxies
(Gunn \& Gott 1972; Biermann 1978; Takeda \etal 1984; Gaetz \etal 1987);
this should be an important effect, and a future paper will compare
stripping with energetic ejection and contrast their effects on the thermal
state of, and metallicity distribution in, the ICM.  Several other
relevant physical processes have been neglected: the heating of the
ICM by galaxy motions and friction (Rephaeli \& Salpeter 1980;
Just \etal 1990); heating by relativistic electrons from radio galaxies (Lea
and Holman 1978; Rephaeli 1979); and thermal and mass transport by
electron conduction  (Rephaeli 1977; David, Hughes, \& Tucker 1992a).
We neglect these effects
because they are considered weak relative to the physics currently
in--place in our simulation algorithm.

\section{III. Models}

\heading{a) Description of Models}

A main purpose of this work is to highlight the differences
in cluster evolution associated with including feedback from
galaxies.  Therefore, in addition to the simulation
of galaxies, gas, and dark matter, we also evolved the same initial
conditions as a two--fluid system of gas and dark matter only.
The version with galaxies then also included ejection
of mass and energy from those galaxies.  In the figures and tables
below, we refer to the run with ejection using the label EJ, while
the run without ejection will be labeled 2F, for two--fluid.

To further make clear the direction in which physical
properties change when feedback is included, we considered a simple
and extreme set of ejection curves.  We assumed that galaxies eject
mass and energy at a flat, constant rate from a redshift $z=4.0$
to the present.  The amplitude of this rate was set by requiring that
over this ejection period, the simulated galaxies blow off half
their mass --- the most extreme scenario considered in simulations
such as David \etal (1991b) and analytical arguments such as White
(1991).  Such a flat mass ejection rate most closely approximates
models in which winds are blown predominantly by Type I supernova
activity (David \etal 1991b).  In setting the wind luminosity, we took
a flat value for a model $10^{10}\lsol$ galaxy and scaled all other
galaxies linearly from this value, assuming a mass to light ratio of
eight.  The value chosen, $4\times10^{42}\ergs$, was a compromise between
the arguments of Bookbinder \etal (1980), whose formalism would suggest
$L_{wind}\,=\, 1-2 \times 10^{43}\ergs$,
and that of White (1991), whose favored values for wind and
galaxy parameters would suggest a luminosity of $8.5 \times 10^{41}\ergs$.
This value is in rough agreement with the mean luminosity of the
half--mass ejection model considered by David \etal (1991b).
Finally, the wind metallicity is set at a flat rate of three times
the solar value.  This value alone is not extreme; other estimates
for the enrichment of the ICM have been based on metal abundances
in winds as high as fifteen times solar (\cf\,Heckman \etal 1990;
Maeder 1990).  Nonetheless, in concert with our other parameters,
this value is enough to overestimate the iron abundance in clusters,
as we will see.

The parameters of the run, and the properties of the specific
set of initial conditions used, are shown in Table I.

\heading{b) Model Evolution}

Figure 1 shows the evolution of a slice through the simulation
volume at six redshifts.  The four columns show the distribution of
dark matter, all gas, ejected gas, and galaxies in the slice; only
$1/7$ of the dark matter and gas particles in the slice are shown
for clarity.  At early times, the filamentary distribution of
matter in the volume is evident.  As the run progresses, matter
drains via these filaments into subclusters, which then collect
to form the final cluster.  In the figure, proto--cluster potential wells
clearly have formed
by a redshift of 1.  Four separate subclusters are in place at a
redshift of a half; these undergo mergers by $z\,=\,0.1$.  Some
elongation from the merger sequence persists by $z\,=\,0.02$, while
the gas has adopted a more rounded distribution since $z\,=\,0.1$.
A smaller group of galaxies, with associated gas and dark matter,
is clear to the north and slightly east of the main cluster,
about $7\mpc$ away.

\heading{c) Temperature, Density, and Metallicity Profiles}

The evolution of the temperature profile for the cluster
is shown in Figure 2, for both the ejection (EJ) and two--fluid
(2F) runs.  Because there exists several comparable subclusters before
$z\,=\,0.246$, we define the ``cluster'' as the most massive of these
in displaying these profiles.  In both cases, the cluster
relaxes to a roughly isothermal distribution within 1--$2\mpc$ at
$z \se 0$.  The inner temperature profile is more nearly level for
the ejection run, though the difference is small.
The temperature decreases
radially as $T \propto r^{-0.88}$ from $\ssim 2 \mpc$ out to about
$7 \mpc$, where it briefly rises because of the subcluster at that
distance (visible in Figure 1).  The isothermal region within 2 Mpc
contains a total mass of $9.4 \times 10^{14} \msol$, corresponding to
a mean interior density $\sim \! 400$ times the mean background value.

The standard descriptive
model for the distribution of matter in clusters
is the so--called hydrostatic isothermal $\beta$--model (Cavaliere
\& Fusco--Femiano 1976, 1978; Sarazin \& Bahcall, 1977).  This
model presumes that fluids making up the cluster are distributed
with a radial density function that can be written as
\beq
\label{eq:betaden}
\rho = \rho_{0} \,
\left\{1 + \left(\frac{r}{r_c}\right)^2\right\}^{-3\alpha / 2}
\eeq
where $\rho_{0}$ is the central density and $r_{c}$ is the ``core
radius.''  In the simplest model, the exponent $\alpha$ for
the gas density profile is equal to $\beta$, defined as
the ratio of specific energies in galaxies and gas,
\beq
\label{eq:betaspec}
\beta \ = \ { \sigma_r^2 \over (kT/\mu m_p) } .
\eeq
Here $\sigma_r$ is the
line--of--sight velocity dispersion of galaxies in the cluster, $T$ is
the gas temperature, $\mu$ the mean molecular weight and $m_p$ the
proton mass.  Beyond the hydrostatic equilibrium assumption, a number
of other assumptions are involved in relating $\beta$
to the ratio of specific energies : (i) the galaxies and binding mass
must have density profiles obeying equation~\ref{eq:betaden} with $\alpha=1$;
(ii) galaxy orbits are isotropic and (iii) the gas is supported purely by
thermal pressure.  Because each of these assumptions are unlikely to be
true to $\sim \! 20\%$ accuracy, the validity of the simple model is
questionable.  In particular, the use of $\beta$ values measured from
X--ray surface brightness profiles as an estimate of the ratio of
specific energies is prone to systematic error (Evrard 1990b; but see
Thomas \& Couchman 1992 for an alternative view).  Still, we
can take equation~\ref{eq:betaden} in the spirit of a fitting function
and then examine the correspondence between the values of $\alpha$ fit and the
ratio of specific energies.  This will be discussed in the following
section.

We fit this model to the actual dark matter
and gas radial density profiles using the average density in
concentric spherical shells centered on the most bound dark matter
particle in the cluster (effectively, the bottom of the potential
well).  The top row of  Figure 3 shows the dark matter density profile,
at $z\,=\,0.02$, for the ejection and two--fluid runs.  The lines
correspond to best--fits for different values of the cutoff radius
--- the maximum radius of data to use in the fit; Table 2 shows the
best--fit parameters.
The dark matter displays very similar forms in the EJ and 2F runs, as
naively expected.  The core radii for the EJ fits are not much more than
our resolution limit; there is no evidence for a core to the dark
matter distribution, in agreement with Dubinski \& Carlberg (1991),
who examined a halo of smaller mass with higher resolution.  Indeed,
the use of equation~\ref{eq:betaden} to model the dark matter is not
well motivated,
since there is no obvious physics operating on the dark matter
from which one could attempt to derive a core radius (Crone, Evrard \&
Richstone 1993).

The central density of the dark
matter in the EJ run is slightly lower than that of the 2F run, and
the core radii and $\alpha_{DM}$ value larger, respectively.
This may be expected if dynamical friction were transferring energy
from the galaxies to the dark matter in the EJ run (recall there are
no galaxies in the 2F model).  Within about $200 \kpc$, the
timescale for orbital decay due to dynamical friction (Binney
and Tremaine 1987) is $\sim \! 1\,$Gyr, so it is not implausible that the
dark matter may be affected in the central regions.  However, the effect
is weak; a second fit to the 2F run, with a $1.0 \mpc$ cutoff and a fixed
$180 \kpc$ core radius (the same as the EJ run for a $1.0 \mpc$ cutoff),
produced values for the central density and $\alpha_{DM}$ equivalent
to those for the EJ run.  Furthermore, the two innermost data points for
both runs, which seem to show the difference in central density, lie below
or near our resolution limit; our resolution is not sufficient to make a
defensible statement on the matter.

It is also worthwhile to note the dependence of the best--fit
parameters upon the cutoff radius.  The dispersion of the data points
at larger radii seems to increase, and there are regions where the
local slope seems to be steeper or shallower than neighboring regions.
This behavior can be expected simply because of the elongated
shape of the dark matter, evident in Figure 1.
The best--fit value of $\alpha_{DM}$ thus is strongly influenced by the
choice of cutoff radius.  It is thus inappropriate to consider
the dark matter density profile of this cluster at large radii to be
a simple power law, $\rho \propto r^{-3 \alpha_{DM}}$; the logarithmic
slope changes with radius.

The gas density profiles of the two runs are compared in the middle
row of Figure 3.  The curves shown are fits to equation (5) for
the same three values of the fitting cutoff radius; Table 3 shows
the best--fit parameters.  The fitted value of $\alpha$ shows a
dependence on the cutoff radius of the fit, just as the dark
matter does.  The central gas density in the EJ run is
lower that that of the 2F run.
This difference is due to the energy input from gas ejection.
Early on, when the local intergalactic medium is still cold,
gas which is blown out of cluster galaxies early drives the
surrounding gas onto a higher adiabat.  Figure 4 displays the entropy
as a function of radius at several redshifts, verifying that the
central gas in the EJ run lies on a higher adiabat than the
gas in the 2F run.  Because the entropy
profile of the cluster rises with radius, the preheated gas
ends up further from the center of the cluster potential
well.  The end result is the ICM being ``puffed up,'' to a lower
central density than a pure infall scenario.
Previously, Kaiser (1991) and Evrard \& Henry (1991) examined
analytic models invoking this effect, while  Navarro, Frenk \& White
(1993) and Evrard (1990b) have run simulations with gas
initially set to a fixed, high entropy.

The corresponding best fit
$\beta$  is smaller in the EJ case than in the 2F run at $z \se 0.02$,
but the effect
is only marginally significant, since $\beta \se 0.83 \pm 0.09$ and
$0.92 \pm 0.08$ in the EJ and 2F models, respectively.  We show below
that the value of $\beta$ deduced from surface brightness profiles changes
with time as the cluster experiences periods of different dynamical
activity (\eg, pre-- or post--merger states).

The bottom row of Figure 3 shows the galaxy density profile of the final
cluster at $z\,=\,9.0$ (the initial distribution of galaxies) and
$z\,=\,0.02$.  The cluster's ``initial location'' was defined by
the center of mass of all the dark matter particles that end up in
the cluster at the end of the run.  The profile steepens and
can be fit to equation~\ref{eq:betaden} with $\alpha\,=\,0.99\pm0.35$
at $z\,=\,0.02$; the number statistics are poor and the
error bars correspondingly larger than for the dark matter and gas
distributions.  The steeper galaxy density profile is in agreement
with the fact that the galaxy population has a lower velocity
dispersion (\ie, is cooler) than the dark matter in the cluster,
as discussed in the following section.  It is also in agreement with
observations.  The ratio of mass in gas to mass in cluster galaxies, in
the ejection run, is less than one within $300\kpc$, and rises steeply
beyond to a value of $5.9$ at an Abell radius.  This is comparable to
the value of $5.1$ inferred for the Coma cluster within that radius
(White \etal 1993).

The radial distribution of ejected gas in the cluster can be
used to define an iron abundance profile.  Figure 5 shows the true
iron abundance profile of the intracluster medium at four redshifts.
The profiles for $z\,=\,0.499$ and $z\,=\,1.01$ were obtained from
the most massive progenitor of the final cluster.  The jagged nature
of the profiles in the center is due to discreteness noise.
A clear metallicity gradient is present, even at a redshift of one.
At all but the earliest redshift shown, the innermost bin has an iron
abundance approaching three times solar; most or all the gas at small
radii is ejected matter.  The average abundance for all gas
within $318\kpc$ (about a core radius) is 1.62 solar; one must go
beyond 2 Mpc for the average abundance within to drop below 0.5 solar.

How could the iron abundance be brought into line
with the observed range of 0.25 -- 0.5 solar for rich clusters?
The abundance distribution could be
decreased in three ways:  by decreasing the metallicity of the ejecta;
by decreasing the amount of ejected matter;
and by adopting a time--varying ejection rate and ejecta iron mass fraction.
Most probably, some combination of the latter two is in order.

It is doubtful that reducing the average metallicity of ejecta is
appropriate; Heckman \etal (1990) argue that starburst--driven galactic
superwinds could in principle be made up by as much as 50\% metals.
However, we considered a model in which galaxies eject half their
initial mass; this is the most extreme case considered by other
studies (David \etal 1991b; White 1991).  Models with smaller ejected
mass fractions are more common.  The ejected mass fraction also
depends on the depth of the galactic potential well; high
mass galaxies eject a lower fraction than low mass galaxies.
Also, energetic
ejection with a flat ejection rate is comparable to the late--time
behavior of galaxies, with supernova activity dominated by Type I
supernovae.  Young galaxies, however, should pass through a phase in
which supernova activity is dominated by Type II supernovae; during such
a phase, the ejection rate and metallicity of ejecta should differ from
late times.  It is also unclear as to whether ejecta from Type I or Type
II supernovae have dominated the enrichment of the ICM, and thus the
quality of our flat rate approximation is unclear.  Finally, some of
the enriched gas may cool to temperatures out of the X--ray region.
In the EJ model, the central cooling time at the end of the run
is longer than a Hubble time; a large--scale cooling flow is unlikely.
However, a higher spatial resolution experiment explicitly including
cooling is needed to address this issue correctly.

Recall that changing the metallicity of the ejecta in our
simulation amounts only to rescaling
the vertical axis of Figure 5 by a constant factor.  Similarly,
changing the total amount of mass ejected --- but
keeping a flat ejection rate --- will only change the fraction of the ICM
made up of ejected matter by a constant factor.  In other words,
so long as the ejection rate and metallicity of ejecta are kept uniform
in time, the {\it shape} of the true iron abundance profile should be the
same in this simulated cluster; at z=0.02, the iron abundance at one core
radius should be $\sim$ 3 times its value at three core radii.  This suggests
that observational evidence for central concentration of iron is
not enough to convincingly argue that ram pressure stripping of
cluster galaxies was the mechanism for its deposition; it is possible
to obtain strong gradients through energetic ejection.

Why does the gradient exist?  In Figure 3, the galaxy number
density profile at the beginning of the simulation shows a shallow
gradient from the comoving location of the center of the final cluster;
this shallow gradient steepens by $z\,=\,0.02$, when the fall-off in
galaxy number density is faster than the fall-off of intracluster gas
density.  Because the galaxies are the source of ejected matter, it seems
reasonable to expect that the more steep galaxy density profile (relative
to the intracluster gas) would lead to a metallicity gradient.
There are certainly other processes
which could play a role.  For instance, since the gas is ejected
at roughly constant temperature in this model, one could envision
ejected gas which was cooler or hotter than its surroundings moving
inward or outward to a region of comparable entropy.  However,
the steepening of the density profile seems a strong enough
effect to easily account for the gradient.  Figure 6 shows the
profiles of three ratios:  a) the mass in galaxies within a radius
{\it r} over the primordial gas mass within {\it r}, at $z\,=\,9.0$;
b) the same quantity at $z\,=\,0.02$; and c) the the mass in ejected gas
within {\it r} over the primordial gas mass within {\it r}, at $z\,=\,0.02$.
These ratios are profiled against the total mass enclosed within {\it r}.
In the absence of mixing between shells of constant mass enclosed
and any strong radial motion of the ejected matter, one would expect
the shape of the profile for c) to be similar to that of b).
The metallicity profile in Figure 5 appears to begin levelling off
at about $1\mpc$ for $z\,=\,0.02$; that corresponds to an enclosed
total mass of about $4.66\times 10^{14}\msol$.  Within that
mass shell, b) and c) appear to have similar shapes.

To summarize, the metal gradient appears to be a consequence of
ejection which takes place while galaxies have a gradient in
number density with respect to primordial gas.  The latter is due
to the fact that the ICM is slightly hotter and
therefore more extended than the galaxy distribution.

The two rich clusters with data on abundance gradients are
Coma and Perseus.  Observations have suggested that Coma has
a fairly flat iron abundance profile
(Hughes, Gorenstein, \& Fabricant 1988; Hughes \etal 1992; Watt \etal 1992).
Imaging of the Perseus cluster by {\it Spacelab 2} (Ponman \etal 1990) and
{\it Spartan--1} (Kowalski \etal 1993) suggest a strong abundance
gradient, decreasing with radius; however, analyses of BBXRT data
disagree with this result (Mushotzky 1991).  It has been suggested
(\eg\/ Watt \etal 1992) that a lack of an abundance gradient in Coma,
compared with a strong gradient in Perseus, is indicative of a difference
in formation history; clusters that formed recently via major mergers
would have any pre--existing abundance gradients ``washed out''
in the merger.  Such mixing after mergers was examined for collisionless
fluids such as stellar systems by White (1980), who found mixing
to be relatively inefficient.  We examine this hypothesis for the
collisional intracluster gas by looking at the iron abundance profiles
of the objects that merge to form the final cluster, and the merger
product, over a time period around the merger.  The abundance profiles
for the three subclusters that merge between $z\,=\,0.499$ and
$z\,=\,0.246$ (see Figure 1), along with their evolution over a period
of $1.47$ Gyr, are shown in Figure 7.  Lines depicting an object
disappear when they merge into the main cluster.  A strong gradient in the
most massive object is clearly in place by a redshift of 0.401; a subsequent
merger with the next most massive subcluster does not seem to diminish
that gradient.  The second merger also seems to have little effect.
We therefore can claim to see no clear evidence for such a relationship
between abundance profile and recent dynamical history in our simulation.

However, it should be noted that the profiles displayed
in Figures 5 and 7 are true mass--weighted radial profiles, based on
the average abundance in concentric spherical shells centered on the
minimum in the potential.  Observers do not have access to this information;
observed profiles are necessarily projected and flux--weighted.
Furthermore, they are routinely centered on or taken from the
x--ray surface brightness maximum rather than the potential
minimum, which is unknown.  While the cluster gas attempts to relax
after mergers, the surface brightness maximum, the iron abundance
maximum, and the potential minimum are often in {\it three different
projected locations}.  This is the source of the ``weaker gradient''
at $z\,=\,0.318$; the profile is centered on the potential minimum,
which is far from the abundance maximum, so the abundance maximum
gets averaged over a large spherical shell and doesn't show up
in the profile.  As the surface brightness and iron abundance
maxima are near each other at that redshift, a flux--weighted
profile centered on the surface brightness maximum continues
to show the strong gradient.  While the offsets between the
two maxima, and the potential minimum, complicate the definition
of a ``central'' iron abundance, this effect does not appear
here to be able to make Coma appear to have a flat abundance profile.
The two maxima are always near enough to each other that a
value of the flux--weighted iron abundance at the surface
brightness peak is still large, even if not the largest; values
at 20' away are generally depressed relative to that central value.

Alternately, the apparatus used to generate X--ray images and surface
brightness profiles (described below) can also be used to create a projected,
flux--weighted metallicity map.  This map indicates the iron abundance
an observer would measure if the observer could point a telescope at every
location in the map.  Sampling the map at various points then creates
abundance profiles more akin to those from real
observations.

Figure 8 shows the results of this process, for the cluster at
$z\,=\,0.020$.  The upper left hand corner is
a flux--weighted metallicity map, using emission in the
ROSAT passband, with a perfect resolution of $0^\prime .5$
($0^\prime .5$ pixel size and
no smoothing with other pixels).  The central concentration of metals is
evident, but also evident is a significant amount of structure in the metal
distribution which might not be expected simply from the x--ray image
(Figure 10).  To the right are four ``profiles'' of the iron abundance
centered on the x--ray surface brightness maximum and measuring the
abundance at $5^\prime$ intervals.  These profiles show the central maximum and
suggest the structure at larger radii.  Note that flux--weighting along
the line of sight has diminished the observed central iron abundance to
about twice solar.

The plot in the lower left hand corner shows a flux--weighted
metallicity map using emission in the passband of
$2\,-\,10\kev$, the same as EXOSAT and HEAO--1.  The pixels in
this map were generated by flux--weighting along the
line of sight assuming an instrument response modelled by a
Gaussian filter with a FWHM of $45^\prime$.
To the right are profiles generated in the same
fashion as above.  Note the different contour levels used from the
above map.  After smoothing, the central iron abundance has
dropped to $\sim \! 1.1$ solar and the structure in the
projected metallicity evident above has vanished.  Also, the
apparent abundance maximum is now over $10^\prime$ from the ``ROSAT''
x--ray surface brightness maximum.  Finally, the contour lines
and profiles make clear that with this heavy smoothing, the
iron abundance profile can appear quite flat, when in reality
a strong gradient exists.

\heading{d) X--ray Characteristics}

To study the X--ray emission that would be expected from
our simulated clusters, we integrated a band--limited thermal
bremsstrahlung emissivity over the volume of interest by using
the equations of SPH to turn the volume integral into a sum over
particles (see Appendix).

Figure 9 shows the (appropriately redshifted) band-limited
X--ray luminosity vs. redshift for the ejection and two--fluid runs,
as a function of redshift.  At high redshift, the luminosity from the
two--fluid run is as much as a factor of five greater than for the
ejection run; it remains roughly constant until $z\approx 0.5$.
The slow accumulation of matter during this time period does not
significantly affect the X--ray luminosity.  After $z\approx 0.5$,
the luminosity climbs sharply, the result of a large merger in--progress.
Both of the peaks/valleys at lower redshifts correspond to merger events.
Again, similar to Evrard (1990b), we see that the X--ray luminosity can
decrease after mergers due to oscillations in the central
density as the cluster attempts to relax.  For the ejection run,
$L_{x}\,=\,0.19$ (in units of $10^{44} \ergs$) at $z\,=\,1.01$; this
climbs by a factor of $\ssim 6.5$ to $1.25$ at $z\,=\,0.02$.  The
two--fluid run climbs from $L_{x}\,=\,1.05$ at $z\,=\,1.01$ to
$2.39$ at $z\,=\,0.02$, an increase by a factor of 2.3.  The
larger increase in the EJ run is a consequence of the fact that the
`preheating' by gas ejection affects primarily the central, low
adiabat gas.  At early times, preheating drives gas onto higher
adiabats, and the emission from the core of the cluster is necessarily
diminished.  As gas piles onto the cluster at larger radii and higher
adiabats, the relative importance of the core emission diminishes,
and the difference between the EJ and 2F runs decreases.  Energetic
winds would thus seem to imply lower luminosities at high redshift
($z \ssim 1$), combined with more rapid evolution to $z \ssim 0$, than
otherwise would take place in hierarchical models.  Rapid evolution
in X--ray luminosity has been hinted at in the observations of
Castander \etal (1993).  Unfortunatly, the small number of clusters in
their sample (5) makes it difficult to conclusively rule out models.
However, the fact that they detected emission from 3 of the clusters
at $z \ssim 1$ should provide an upper limit to the strength
of feedback from galactic winds, since extremely strong input would
make the clusters invisible at high redshift.
We intend to examine this issue in a future paper.

Figure 10a shows a sequence of simulated ROSAT PSPC images
for the run with galaxies and ejection, while Figure 10b shows
corresponding images for the two--fluid run.  The evolution of
the model proceeds clockwise from the upper left.  The outer
surface brightness contour, and the spacing between contours,
are the same for both figures and throughout each figure.  The
black dots in the run with galaxies/ejection corresponds to the
projected locations of the galaxies in the simulation.  The
calculated incident flux in the $0.1-2.4 \kev$ band
was converted to ROSAT PSPC counts per second
using a simple conversion factor taken to be
(Rosat Scientific Data Center, 1989)
\beq
\label{eq:ROSAT}
1.4 \pspc \ctss \,=\,1 \times 10^{-11}\cgsflux,
\eeq
for a thermal spectrum in this temperature range.  The ``exposure''
was simulated as 7.2 ksec in length.
Because of the preheating, there are sources of X--ray emission
in the 2F run visible at high redshift that are too dim to be seen
in the run with ejection.  At lower redshifts, the size
of the main object is roughly comparable in the two runs, but there
are more contours in the 2F model.  The ejection run achieves
higher cluster temperatures, but lower densities, and thermal
bremsstrahlung emission is more strongly sensitive to the density
of the gas.

The stronger emission shown in Figure 10b allows us to see
the mergers implied in the luminosity evolution (Figure 9).  Of the
four central sources of emission at $z\approx0.5$, the three northern
sources merge between $z\,=\,0.5$ and $z\,=\,0.3$; this is the source
of the first peak in Figure 9, and the strong central source at
$z \approx 0.25$ in Figure 10b.  The source to the south merges at
$z \approx 0.1$, but still retains some dynamical identity.  Thus,
in addition to the second peak on Figure 9, we see the $z\,=\,0.102$
isophotal contours in Figure 10b as stretched to the southeast.
Momentum from this sub--lump subsequently drives north, causing
the contour lines to be stretched in that direction at later redshifts
(\eg, Roettiger, Burns, \& Loken 1993).

The images shown have been fit to a radially--symmetric
surface brightness profile of the usual form implied by the
$\beta$--model,
\beq
\label{eq:sb}
\Sigma_{x}\left(\theta\right) \,=\, \Sigma_{0}
	\left[ 1\,+\,\left(\theta / \theta_{x}\right)^{2}\right]
	^{-3\beta_{fit}\,+\,1/2}.
\eeq
Poisson noise was added at an amplitude of $3\times 10^{-4} \sbunits$;
a flat background of this value was then subtracted off.  Only
bins with surface brightnesses higher than this value
were used in the fits.  Figure 11 shows the values and errors for the
surface brightness in each radial bin, as well as the best fit, for
the four lowest redshifts imaged in each run.
Table 4 shows the best fit values
and errors for the central surface brightness $\Sigma_{0}$, X--ray
core radius $r_{x}$ (for
$\theta_{x}$ at redshift z), and exponent $\beta_{fit}$, for the
eight fits.  In all cases, Equation~\ref{eq:sb} provides
a good fit to the surface brightness profiles.  At redshifts
$z \lta 0.1$, the values of the fitting parameters
are comparable to those obtained in rich cluster observations.

The values of $\beta$ at $z \se 0.02$ derived from imaging are
consistent with, but not exactly equal to, the value obtained above
from directly fitting the $3D$ density profile.  The errors are
smaller for the surface brightness measurement because they are based
on Poisson statitics of photon counts.  The difference is not
surprising given the level of asymmetry in the X--ray profiles.
The $\beta_{fit}$ value of $0.74 \pm 0.01$  for the EJ run is again
smaller than the $0.79 \pm 0.01$ value for the 2F run.  However, this
ranking does not hold at all times.  The previous output frame at $z
\se 0.06$ shows the EJ profile is steeper than the 2F profile.
Evidently, the effect on $\beta_{fit}$ due to feedback is not larger
than the variance caused by dynamical effects.

\section{IV. Mass Estimates, the Cluster Baryon Fraction and
the $\beta$--Problem }

\heading{a) X--ray and Optical Mass Estimates}

Under the $\beta$--model, it is assumed that the gas density is
well--described by Equation~\ref{eq:betaden}, with $\beta$ and $r_c$
obtained from the cluster's X--ray surface brightness
profile.  Furthermore, by assuming the gas is isothermal, the central
surface brightness can be simply related to the emission measure along a
line of sight through the center of the cluster, and thus to the central
gas density.  This leads to a relation between central number density
$n_{0}$ and X--ray observables of the form
\begin{eqnarray}
\lefteqn{ n_{0} \,=\, 1.34 \times 10^{-3} \cc \cdot
\frac{\left(1\,+\,z\right)^{2}}{\left(f_{band}\left(T,z\right)\right)^{1/2}}
\cdot \nonumber
\frac{\Gamma\left(3\beta\right)}{\Gamma\left(3\beta \,-\, 1/2\right)} } \\
 & & \cdot
\left(\frac{\Sigma_{0}}{10^{-2} \sbunits}\right)^{1/2}
\cdot
\left(\frac{T}{10^{8}\K}\right)^{-1/2}
\cdot
\left(\frac{r_{c}}{250 \kpc}\right)^{-1/2},
\end{eqnarray}
where we have figured in our ROSAT response approximation,
Equation~\ref{eq:ROSAT}; and $f_{band}\left(T,z\right)$ is the fraction of
emission that falls into the the ROSAT passband (see Appendix).
Thus, $\Sigma_{0}$,
$\beta$, $r_c$, and the cluster temperature give the central density $n_{0}$;
integrating the density above gives the gas mass within a certain radius.

Assuming that the ICM is both hydrostatic and
ideal, the total mass within a given radius is given by
\beq
\label{eq:bindmass1}
M\left(<r\right) \,=\, \frac{k T\left(r\right) r}{\mu m_{p} G}
	\left[ - \frac{d\,\ln \rho_{g}}{d\,\ln r}
	- \frac{d\,\ln T}{d\,\ln r}\right].
\eeq
If the gas is isothermal and the density follows Equation~\ref{eq:betaden},
we have
\beq
M_{bind}\left(<r\right) \,=\, \frac{k T\left(r\right) r}{\mu m_{p} G}
	\cdot 3\beta
	\cdot \frac{\left(r/r_{c}\right)^2}{1\,+\,\left(r/r_{c}\right)^2}.
\eeq
For large radii ($r \gg r_{c}$), the last fractional factor goes to 1 and
can be ignored; but for smaller radii, ignoring this factor will result
in overestimates of the binding mass.  At $r \,=\, 3r_{c}$, for instance,
a 10\% error is introduced.

To compare the results of the simulation with such a binding
mass estimate, an additional correction to this equation is necessary.
Equation~\ref{eq:bindmass1} was derived assuming the usual hydrostatic
equation, which in turn assumes an $r^{-2}$ force law.  However,
the force law used in the simulations comes from the softened
point--mass potential, $V(r)\,\propto\,(r^{2}\,+\,\epsilon^{2})^{-1/2}$,
and not a Coulomb law.  Furthermore, for such a force, Gauss' law
no longer holds; the importance of the correction depends on the distance
away from the cluster center.  Thus, the hydrostatic equation becomes
\beq
\frac{\vec{\nabla}P}{\rho} \,=\, -\,S\left(r/\epsilon\right) \cdot
\frac{GMr}{\left(r^{2}\,+\,\epsilon^{2}\right)^{3/2}}\hat{r},
\eeq
where $S$ is a function that depends on an integral over
the mass distribution.  Our densities at large radii are roughly
proportional to $r^{-2}$, and with such a distribution, $S$
is approximately unity.  The formula
for estimating binding masses in our simulations is then
\beq
\label{eq:bindmass2}
M\left(<r\right) \,=\, \frac{k T\left(r\right) r}{\mu m_{p} G}
        \cdot 3\beta
        \cdot \frac{\left(r/r_{c}\right)^2}{1\,+\,\left(r/r_{c}\right)^2}
        \cdot \left[1\,+\,\left(\epsilon/r\right)^2 \right]^{3/2}.
\eeq

Figure 12 shows the actual gas mass (boxes) and total mass
(crosses) contained within, for the ejection run at a redshift of
$z=0.02$.  Also shown for comparison are predictions for the
enclosed gas and total masses, described above.  The
values of $\beta$, $r_{c}$ (or $r_{x}$), and $\Sigma_{0}$ (thus $n_{0}$)
are from the corresponding z--axis projection surface brightness
fit (see Figure 11 and Table 4).  Similar results are found for the
other two projections.  The temperature used in the
binding mass estimate is the ``central temperature'' at that
redshift from the true profile discussed in c) above.  Both
the inferred gas and binding masses are accurate to better than $20\%$
beyond $500 \kpc$.  There is a tendency to underestimate
the binding mass which grows larger at smaller radii.  However, at
scales above our resolution limit, the error in this mass estimate is
never larger than $\sim 45\%$, comparable to the error found in
two--fluid simulations (Navarro \etal 1993;
Evrard, Metzler \& Navarro 1993).  The introduction of the additional
physics of galactic winds does not detrimentally affect the accuracy
of mass estimates drawn from the $\beta$--model.

Optically, one can obtain a binding mass estimate using
the first velocity moment of the collisionless Boltzmann equation.
The ``hydrostatic equation'' for galaxies can be written as
\beq
M\left(<r\right)\,=\,
-\frac{r^{2}}{G\,\rho_{gal}}
\frac{d\left(\rho_{gal}\sigma_{r}^{2}\right)}{dr}
-
\frac{2r\sigma_{r}^{2}A\left(r\right)}{G},
\eeq
where $\sigma_{r}$ is the radial velocity dispersion, and
$A\left(r\right)\,=\,1\,-\,\sigma_{t}^2/\sigma_{r}^2$ is the
anisotropy parameter.  If we further assume that
cluster velocity dispersions are isotropic ($A\left(r\right)\,=\,0$)
and independent of radius, we get
\beq
M\left(<r\right)\,=\,\frac{r \sigma_{r}^{2}}{G}
\left(-\frac{d\,\ln \rho_{gal}}{d\,\ln r}\right).
\eeq
Assuming $\rho_{gal}$ is well fit by Equation~\ref{eq:betaden},
and fitting the galaxy density profile out to 3 Mpc, the logarithmic
derivative
$-d\,\ln \rho_{gal}\;/\;d\,\ln r\,\ssim\,3\alpha_{GAL}\,=\,2.7$.
If we plug in the line--of--sight velocity
dispersions of 606, 960, and 655 $\kms$ obtained by viewing the
cluster along the simulation's principal axes (drawn from
61, 94, and 64 galaxies within the projected Abell radius), we get
mass estimates (in units of $10^{15}\msol$) of $0.69$, $1.7$ and $0.81$
respectively.   The actual mass within an Abell radius is
$1.2\times10^{15}\msol$.
The inaccuracy is due to error in our assumptions of isotropy or
radially--constant velocity dispersions.  Additionally, for one
projection, the infalling subcluster apparent in Figure 1
falls within the projected 3 Mpc radius; a large number of
infalling objects boosts the sample size but artificially
inflates the estimate of velocity dispersion.  Interestingly, if
we abandon the isotropy assumption and avoid projection effects
by using an average 1--D velocity dispersion determined from
real 3--D data of $769 \kms$, we obtain an estimate of
$M\left(<\,3\mpc\right)\,=\,1.2\times10^{15}\msol$ equal to
the true value.   Again, however, this agreement is fortuitous
and, at any rate, impossible to measure observationally.

Another method of estimating virial masses is by use of a projected
estimator, such as Equation 10--23 in Binney and Tremaine (1987):
\beq
\label{eq:bntmass}
M_{vir}\,=\,\frac{3 \pi N^2 \sigma^{2}}{2 G}\,
\frac{1}{\sum_{i=1}^{N}\;\sum_{j<i}|\vec{R_{i}} - \vec{R_{j}}|^{-1}},
\eeq
with $N$ the number of galaxies, $|\vec{R_{i}} - \vec{R_{j}}|$
the projected separation between galaxies $i$ and $j$, and
$\sigma$ the one--dimensional velocity dispersion as usual.
Applying this estimator to our cluster along the same three
axes yields mass estimates (in units of $10^{15}\msol$) of $0.49$,
$1.3$, and $0.58$ respectively.
This technique is known to underestimate binding masses where
the dark halo is more extended than the galaxy distribution, as is
the case here.  For the second axis, this underestimate
is countered by the overestimate of $\sigma$ due to
projection effects from the infalling subcluster,
resulting in a mass estimate which is actually close to the mark,
but is nonetheless spurious.

\heading{b) The Cluster Baryon Fraction}

The observed baryon fraction of Coma and other rich clusters appears
to exceed the limits on the global baryon fraction set by nucleosynthesis
constraints if $\Omega \se 1$ (White \etal 1993).  Above, we described how
one obtains estimates of
the gas and binding mass within a certain radius from X--ray observations.
Galaxy counts and luminosities, combined with a
reasonable mass--to--light ratio, allow an estimate of the mass
in cluster galaxies $M_{gal}$.  With these three quantities, one can estimate
the baryonic mass fraction of the cluster
\beq
f_{b} = \frac{M_{gal} + M_{ICM}}{M_{tot}}
\eeq

In the absence of dissipation, one would expect the baryon
fraction in the cluster to be representative of the baryon fraction
of the universe; the baryons and dark matter in a cluster fell in
from roughly the same comoving volume.  We then have
\beq
f_{b} = \frac{\rho_{b}}{\rho_o} = \frac{\Omega_{b}}{\Omega_o}.
\eeq
Cluster mass estimates give us $f_{b}$ and big bang nucleosynthesis theory
gives limits on $\Omega_{b}$.  The combination yields a prediction for
the cosmological density parameter $\Omega_o$.

It is interesting
to ask what baryon fraction would be inferred from `observing' the
simulated cluster.  Figure 13 shows the true baryon fraction
as a function of radius for the runs with and without
ejection.  Also shown are an observer's {\it prediction} for this
function; $M_{ICM}$ and $M_{tot}$ are taken from the hydrostatic isothermal
$\beta$--model mass estimates shown earlier, while $M_{gal}$ is
determined by assuming that the observer knows perfectly the three-space
distribution of galaxies (not just the projected distribution) as
well as their actual masses.  The true value of the baryon fraction for
the simulation, $f_{b}\,=\,0.1$, is shown as a dotted line across the plots.
The central baryon fraction is above the global value in the ejection run
because of the introduction of galaxies; baryons that may not have made it
deep into the cluster in the 2F run are carried in via the collisionless
galaxy particles.  The predicted value generally lies above the true
value at low radii because the hydrostatic mass estimator is
underestimating the binding mass at these radii, as shown in Figure
12.  Beyond the central region, the predicted baryon fraction for both runs
is never off by more than $50\%$ from the true value.  At 1 Mpc, the
predicted value in both runs is similar to the global value; vigorous winds
do not seem to dramatically affect the accuracy of estimates of the baryon
fraction.

The best value of $\Omega_{b}$, from comparison of nucleosynthesis
calculations with observed elemental abundances, is
$\Omega_{b}\,h_{50}^{2} = 0.05 \pm 0.01$ (Walker \etal 1991).
Meanwhile, observations of rich clusters suggest baryon fractions of
15\% -- 30\% (\eg Briel, Henry, \& B\"{o}hringer 1992; David \etal 1992b).
The two can be reconciled if the universe is open $\Omega_{o} = 0.15-0.3$.
Analysis of our simulated cluster, which has an X--ray and optical
appearance fairly typical of observed clusters, indicates that
observational estimates of the baryon fraction are accurate to within
$\sim \! 50\%$.  Reconciling the cluster baryon fraction with $\Omega
\se 1$ is not possible solely by appealing to large systematic errors
in mass estimates.

\heading{c) Velocity bias and the $\beta$--problem}

Although they reside in a common potential well,
the specific energies of the different mass components of the cluster
---- dark matter, galaxies and gas --- need not be exactly equal.
Energy can be transferred among components; for example, by dynamical
friction of dark matter acting on galaxies, or, for the gas, energy
can be stored in either thermal or kinetic form, the latter mainly in
bulk motions of the gas as it attempts to settle to equilibrium in a
time dependent potential well.  Following the $\beta$--model
convention, we define two values of $\beta$ based on the velocity
dispersions of the dark matter and galaxies
$\beta_{DM} \sequiv \sigma_{DM}^2 / (kT/\mu m_p)$ and
$\beta_{gal} \sequiv \sigma_{gal}^2 / (kT/\mu m_p)$.
We use for $T$ a mass weighted measure of the gas
temperature.  Because of the slightly negative radial temperature
gradient, flux weighted temperatures are higher by $3\%$ and $5\%$
in the EJ and 2F runs, respectively, consistent with the results of
Navarro \etal (1993).  We also define a
`velocity bias' parameter (Carlberg 1991) as the ratio of velocity
dispersions in galaxies and dark matter $b_v \sequiv \sigma_{gal}/
\sigma_{DM}$.   All quantities are quoted within a comoving $1 \mpc$
radius from the cluster center.

Figure 14 shows the evolution with redshift of $\beta$
for each species and $b_v$.
No projection effects are considered;
the temperature used is the true mass--averaged gas temperature over
this volume, while the velocity dispersions used are $1\,/\,\sqrt{3}$
of the true 3--D velocity dispersions, calculated from the particle
distributions.  The plot seems to show a periodicity, but this is
deceptive; when time is used as the abscissa, no periodicity emerges.
The source of the jumps in the value of $\beta$ for the
galaxies is the low number statistics involved; no more than thirty or
so galaxies are within the comoving $1 \mpc$ radius at any time, and
three galaxies infalling into the region can sharply change the
velocity dispersion.  After $z \se 0.4$, the value of $\beta_{gal}$
lies in the range $0.5-1.1$, with a time average value of 0.74.

The data show a persistent velocity bias.  The ratio of velocity
dispersions $b_{v}$ climbs to 1 only once, at very high redshift;
most of the time is spent at values between 0.8--0.9.
The time average value over the same period, $<b_{v}>\,=\,0.86$,
implies a modest bias, consistent with results from two---fluid
simulations modelling galaxy formation explicitly
(Evrard, Summers \& Davis 1994).

The presence of velocity bias conflicts with the conclusions of
Lubin and Bahcall (1993), who claim no evidence for velocity
bias in a sample of 41 clusters with both measured X--ray temperatures
and velocity dispersions based on 20 or more redshifts per cluster.
The sample is largely that of David \etal (1993), who compiled
temperatures based on the \einstein Monitor Proportional Counter (MPC),
supplemented with data from {\it Ginga}, {\it EXOSAT} and {\it
HEAO-1}, along with cluster velocity dispersions from the compilation
of Struble \& Rood (1991).  Lubin \& Bahcall (1993) claimed the data
were well fit by a scaling relation $\sigma \spropto T^{0.5}$,
implying a constant $\beta$ with best fit value $\beta \se 0.94 \pm
0.08$.  By assuming the X--ray temperature was an unbiased
estimator of the dark matter velocity dispersion (\ie, $\beta_{DM}
\sequiv 1$), they directly connected $\beta$ to the velocity bias
$\beta \se b_v^2$.  The $\beta$ determination then implies $b_v \se 0.97
\pm 0.04$.  Empirically, there seems to be no room for velocity bias.

However, there are some troubling aspects of this analysis.
The values of $\beta$ span a large range, from $0.28$ (A1142) to $3.13$
(A1775), and the largest values tend to be the least accurately
determined.
Four clusters have values of $\beta$ which lie more than $3\sigma$ away
from the best fit value of $0.94$, and {\it all} lie below the best
fit value.  (The clusters are
A85 ($\beta \se 0.55 \pm 0.10$), A754 ($\beta \se 0.51 \pm 0.12$),
A1060 ($\beta \se 0.57 \pm 0.12$) and A1656
(Coma, $\beta \se 0.56 \pm 0.10$).)  On the other hand, there are
no clusters which lie more than $3\sigma$ away from $\beta
\se 0.74$, the time--averaged value from our simulation, which includes
a modest velocity bias.  Viewed this way, the observational data could
be interpreted as actually {\it favoring} the presence of velocity bias.

At these lower redshifts, the values of $\beta_{DM}$ lie in the
range $0.9-1.1$, demonstrating that the ICM gas is able to thermalize
rather efficiently.  Still, there is a small amount ($\sim \! 10-15\%$)
of energy in kinetic form. This `loss' is compensated by the energy input
from ejection.  As evidence, the values of $\beta_{DM}$ inferred
from the 2F run are typically $\sim \! 15\%$ above the values in EJ
run.  These values are consistent with values of $\beta$ from other
two--fluid simulations (Evrard 1990a,b; Navarro \etal 1993).
The true values of $\beta$
for each species in the ejection run, as well as for the dark matter
in the two--fluid run, measured within a $1\mpc$ comoving radius from
the center of mass of the cluster, are shown in Table 6.  For the EJ
run, ``observed'' values of $\beta_{spec}$, using the line--of--sight
velocity dispersion from all galaxies within a projected Abell radius
and an emission--weighted central temperature determined from
a mock EXOSAT observation (2--10 keV, 45' FWHM), are $0.41$, $1.03$
and $0.48$ for views along the three principal axes.

\section{IV. Summary}

We have begun investigation of the effects of feedback due to
winds from early--type galaxies on the dynamic and thermal history of
intracluster medium.  Compared to an otherwise identical model without
ejection (a pure infall model), we find feedback affects the structure
of the X--ray emitting gas in a Coma sized cluster in the following ways.

\begin{itemize}
\item[$\bullet$]  At early times, feedback raises the entropy of
proto-cluster gas, preventing it from being compressed to
densities as high as those achieved in the infall case.  Later gas
settling onto higher adiabats is less affected.  This results in both
lower X--ray luminosities and more rapid X--ray evolution in the model
with feedback compared to the infall case.
Statistically complete surveys extending to $z \simeq 1$ would provide
constraining power on the amount of feedback in clusters, although in a
cosmologically dependent fashion.  There is evidence from ROSAT
observations of an optically selected, high redshift sample that
Coma--sized clusters were much dimmmer in X--rays at
$z \simeq 1$ (Castander \etal 1993), but the small sample size
(five clusters) and uncertainties due to projection effects
on optical selection at high redshift hinder serious analysis.
Clusters from the NEP region of the ROSAT all--sky
survey may prove useful; empirical models suggest that roughly 30
clusters with $z > 0.4$ (about $10\%$ of the total) should be found
(Evrard \& Henry 1991).

\item[$\bullet$] The ratio of specific energies in galaxies
and gas, the $\beta$ parameter, is less than one in the ejection
model.  This differs from the infall case where $\beta > 1$ is
generic, as emphasized by Evrard (1990a,b).  The different behavior
arises from the combined effects of a slightly higher gas temperature
(due to energy input from the galactic winds) and a velocity bias
for the galaxy population.  For our Coma--sized model, the temperature
increase is $\ssim 15\%$ over the infall case and the ratio of galaxy
to dark matter velocity dispersions (the velocity bias) is typically
$0.85$.  Since the infall models do not explicitly model galaxies, an
assumption of no velocity bias is usually made, leading to estimates of
$\beta \simeq 1.2$.  Incomplete thermalization of the gas and poor (at
the $\ssim 15\%$ level) modelling of the dark matter distribution are
responsible for $\beta > 1$ in this case (Evrard 1990b).  We favor the
ejection model predictions over the infall case because of the
improvement in physical modelling represented by explicitly including
the galaxy population.  We note that several simulations using
independent methods have produced a value of the velocity bias similar
in magnitude to that seen in our ejection experiment (see Evrard,
Summers \& Davis 1993 and references therein).  Interpretation of
observational data remains problematic.  Edge \& Stewart (1991)
favored spectroscopic values of
$\beta < 1$, but noted that $\beta > 1$ was common for the richest
clusters. Lubin \& Bahcall (1993) claimed the data are consistent with a
constant value of $\beta$ very close to unity, although
individual values span a range from $0.28$ to $3.1$.   A combination of
multi--fiber optical spectroscopy  and spatially resolved X--ray
spectroscopy with the ASCA satellite on a statistical sample of clusters
would go a long way toward settling this matter.

\item[$\bullet$] The morphology of X--ray emission is little affected
by feedback; both the ejection and infall runs produce clusters with
moderately elliptic contours well fit
by the standard form (equation 8) with values of $\beta_{fit}$
in the range $0.7-1$.  There is a slight tendency for lower
$\beta_{fit}$ values in the ejection run, as might be expected from
energetic arguments.  However, this difference is swamped by
variations in $\beta_{fit}$ caused by dynamics.  Mergers are common
and the relaxation time in the outer parts of the cluster is
sufficiently long that the density structure at large radii ($r \gta 1
\mpc$) generally deviates from its hydrostatic equilibrium expectation.
One should not expect tight agreement between $\beta_{fit}$ and
$\beta_{gal}$ in {\it any} model in which clusters formation involves
a large degree of recent merging.

\item[$\bullet$] Estimates of the gas and binding mass profiles of our
simulated clusters, based upon x--ray observations and subsequent
application of the $\beta$--model and
hydrostatic equilibrium, are accurate to better than 20\% for
both the EJ and 2F runs at radii beyond $500 \kpc$.
The inclusion of energetic winds does not seem to affect the
state of cluster gas
strongly enough to detrimentally affect the quality of such
mass estimates.  Dynamical mass estimates based on galaxy
line--of--sight velocity dispersions, on the other hand, showed
errors of as much as factors of 2--3.

\item[$\bullet$] Heavy element enrichment by winds produces a
radially decreasing abundance gradient.
The average iron abundance within $300 \kpc$ is $1.6$ solar,
dropping below $0.5$ solar at radii beyond $2 \mpc$.
The gradient is due to the steeper radial profile
of galaxy number density with respect to intracluster gas.
A central concentration of iron is not a unique signature of
ram pressure stripping.  The metal gradient showed no sign of
weakening during a modest merging encounter.
Observed with a non--imaging,
spectroscopic satellite similar to EXOSAT or HEAO-1, the gradient
becomes extremely difficult to detect, due to contamination by core
emission in  off--center pointings.  Arcmin resolution spectroscopy,
such as is available on ASCA, would uncover the radial gradient.
We predict it should also reveal
small--scale structure in the iron abundance (see Figure 8).
The normalization of the iron abundance
is uncertain by a factor of at least two, reflecting the range of
plausible wind models of early type galaxies.   The strength of the
radial gradient and the degree of small--scale structure are
also likely to be sensitive to details of the wind model employed.

\end{itemize}

There is much work yet to be done before we can claim a true
understanding of the origin and history of the ICM and
its metal content.  The work presented in this paper represents a
first essay at modelling the effect of galactic winds in three
dimensions in a cosmologically viable scenario.  The
nature of the problem is such that there is, unavoidably,
a moderately large parameter space associated with the modelling
which embodies the present uncertainties in galaxy formation and in
modelling winds from evolving galaxies.

We are currently exploring the sensitivity of the results above to
parameters controlling the wind modelling (ejection rates,
luminosities, wind metallicity, etc.).  For a fixed feedback model,
we intend to study a set of clusters spanning a reasonable range of
richness, in order to examine the effect of feedback on correlations
of X--ray and optical properties.  We also intend to study models
in which gas--dynamical stripping is the dominant source of cluster
metals.  Finally, we will pursue higher spatial and mass resolution
experiments which will allow us to examine the effects of
mergers on abundance gradients in more detail, as well as the
effects of cooling on the overall metallicity and abundance gradients.

\section{VI.  Acknowledgements}

\par  CAM would like to thank Hans B\"{o}hringer, Joel Bregman, Larry
David, Rick Mushotzky, Alvio Renzini, and Doug Richstone for very
educational discussions.  This work was supported by NASA Theory Grant
NAGW-2367.

\baselineskip=16.0 truept
\section{References}

\bref
Bardeen, J. M., Bond, J. R., Kaiser, N., and Szalay, S. 1986, \apj,
{\bf 304}, 15.

\bref
Bertschinger, E. 1987, \apjlett, {\bf 323}, L103.

\bref
Biermann, P. 1978, \aa, {\bf 62}, 255.

\bref
Binney, J., and Tremaine, S. 1987, {\it Galactic Dynamics} (Princeton
University Press, Princeton).

\bref
Blumenthal, G. R., Faber, S. M., Primack, J. R., and Rees, M. J. 1984,
\nature, {\bf 311}, 517.

\bref
Blumenthal, G. R., and David, L. P. 1992, \apj, {\bf 389}, 510.

\bref
Bookbinder, J., Cowie, L. L., Krolik, J. H., Ostriker, J. P., and
Rees, M. 1980, \apj, {\bf 237}, 647.

\bref
Briel, U. G., Henry, J. P., and B\"{o}hringer, H. 1992, \aa, {\bf 259}, L31.

\bref
Carlberg, R. G. 1991, \apj, {\bf 367}, 385.

\bref
Carr, B. J., Bond, J. R., and Arnett, W. D. 1984, \apj, {\bf 277}, 445.

\bref
Castander, F. J., Ellis, R. S., Frenk, C. S., Dressler, A., and
Gunn, J. E. 1993, \nature, submitted

\bref
Cavaliere, A., and Fusco--Femiano, R. 1976, \aa, {\bf 49}, 137.

\bref
Cavaliere, A., and Fusco--Femiano, R. 1978, \aa, {\bf 70}, 677.

\bref
Ciotti, L., D'Ercole, A., Pellegrini, S., and Renzini, A. 1991,
\apj, {\bf 376}, 380.

\bref
Cowie, L. L., and Perrenod, S. C. 1978, \apj, {\bf 219}, 354.

\bref
Crone, M. M., Evrard, A. E., and Richstone, D. O. 1993, in preparation.

\bref
David, L. P., Arnaud, K. A., Forman, W., and Jones, C. 1990a, \apj,
{\bf 356}, 32.

\bref
David, L. P., Forman, W., and Jones, C. 1990b, \apj, {\bf 359}, 29.

\bref
David, L. P., Forman, W., and Jones, C. 1991a, \apj, {\bf 369}, 121.

\bref
David, L. P., Forman, W., and Jones, C. 1991b, \apj, {\bf 380}, 39.

\bref
David, L. P., Hughes, J. P., and Tucker, W. H. 1992a, \apj, {\bf 394},
452.

\bref
David, L. P., Jones, C., and Forman, W. 1992b, \baas, {\bf 24}, 1145.

\bref
Davis, M., Efstathiou, G., Frenk, C. S., and White, S. D. M. 1985,
\apj, {\bf 292}, 371.

\bref
Dressler, A. 1980, \apj, {\bf 341}, 26.

\bref
Dubinski, J., and Carlberg, R. G. 1991, \apj, {\bf 378}, 496.

\bref
Edge, A. C., and Stewart, G. C. 1991, \mnras, {\bf 252}, 428.

\bref
Efstathiou, G. and Eastwood, J.W. 1981, {\it M.N.R.A.S.\/}, {\bf 194}, 503.

\bref
Efstathiou, G., Davis, M., Frenk, C. S. and White, S. D. M. 1985,
{\it Ap. J. Suppl.\/}, {\bf 57}, 241.

\bref
Evrard, A. E. 1988, \mnras, {\bf 235}, 911.

\bref
Evrard, A. E. 1990a, {\it Clusters of Galaxies},
edited by W. Oegerle, M. Fitchett, and L. Danly (Cambridge University Press,
Cambridge), 287.

\bref
Evrard, A. E. 1990b, \apj, {\bf 363}, 349.

\bref
Evrard, A. E., Silk, J., and Szalay, A. S. 1990, \apj, {\bf 365}, 13.

\bref
Evrard, A. E., and Henry, J. P. 1991, \apj, {\bf 383}, 95.

\bref
Evrard, A. E., Summers, F. J., and Davis, M. 1994, \apj, in press, 10 Feb.

\bref
Evrard, A. E., Metzler, C.A. \& Navarro, J.N. 1993, in preparation.

\bref
Frenk, C. S., White, S. D. M., Davis, M., and Efstathiou, G. 1988, \apj,
{\bf 327}, 507.

\bref
Gaetz, T. J., Salpeter, E. E., and Shaviv, G. 1987, \apj, {\bf 316}, 530.

\bref
Gingold, R.A., and Monaghan, J. J. 1977, \mnras, {\bf 181}, 375.

\bref
Gingold, R. A., and Perrenod, S. C. 1979, \mnras, {\bf 187}, 371.

\bref
Gull, S. F., and Northover, K. J. 1975, \mnras, {\bf 173}, 585.

\bref
Gunn, J. E., and Gott, J. R. 1972, \apj, {\bf 176}, 1.

\bref
Heckman, T. M., Armus, L., and Miley, G. K. 1987, \aj, {\bf 93}, 276.

\bref
Heckman, T. M., Armus, L., and Miley, G. K. 1990, \apjsup, {\bf 74}, 833.

\bref
Hockney, R.W. and Eastwood, J.W. 1981, {\it Computer Simulation using
Particles\/} (McGraw Hill International, New York).

\bref
Hughes, J. P., Gorenstein, P., and Fabricant, D. 1988, \apj, {\bf 329}, 82.

\bref
Hughes, J. P., Butcher, J. A., Stewart, G. C., and Tanaka, Y. 1992, \apj,
{\bf 404}, 611.

\bref
Jones, C., and Forman, W. 1992, {\it Clusters and Superclusters of Galaxies},
edited by A. C. Fabian (Kluwer Academic Publishers, Dordrecht), 49.

\bref
Just, A., Deiss, B. M., Kegel, W. H., Bohringer, H., and Morfill, G. E. 1990,
\apj, {\bf 354}, 400.

\bref
Kaiser, N. 1991, \apj, {\bf 383}, 104.

\bref
Katz, N., and White, S. D. M. 1993, \apj, {\bf 412}, 455.

\bref
Katz, N., Quinn, T., and Gelb, J. 1993, \mnras, in press.

\bref
Kowalski, M. P., Cruddace, R. G., Snyder, W. A., Fritz, G. G.,
Ulmer, M. P., and Fenimore, E. E. 1993, \apj, {\bf 412}, 489.

\bref
Larson, R. B., and Dinerstein, H. L. 1975, \pasp, {\bf 87}, 911.

\bref
Lea, S. M. 1976, \apj, {\bf 203}, 569.

\bref
Lea, S. M., and Holman, G. D. 1978, \apj, {\bf 222}, 29.

\bref
Loveday, J., Peterson, B. A., Efstathiou, G., and Maddox, S. J. 1992,
\apj, {\bf 390}, 338.

\bref
Lubin, L. M., and Bahcall, N. A. 1993, \apjlett, submitted.

\bref
Maeder, A. 1990, in {\it Massive Stars in Starbursts},
edited by C. Leitherer, T. Heckman, C. Norman, and N. Walborn
(Cambridge University Press, Cambridge).

\bref
Matteucci, F., and Tornamb\`{e}, A. 1987, \aa, {\bf 185}, 51.

\bref
Matteucci, F., and Vettolani, G. 1988, \aa, {\bf 202}, 21.

\bref
McCarthy, P., Heckman, T., and van Bruegel, W. 1987, \aj, {\bf 93}, 264.

\bref
Monaghan, J. J. 1992, \annrev, {\bf 30}, 543.

\bref
Mushotzky, R. F. 1991, {\it Clusters and Superclusters of Galaxies},
edited by A. C. Fabian (Kluwer Academic Publishers, Dordrecht), 91.

\bref
Navarro, J.N., Frenk, C.S., \& White, S.D.M. 1993, preprint.

\bref
Perrenod, S.C. 1978, \apj, {\bf 226}, 566.

\bref
Ponman, T. J., Bertram, D., Church, M. J., Eyles, C. J., Skinner, G. K.,
Watt, M. P., and Willmore, A.P. 1990, \nature, {\bf 347}, 450.

\bref
Rephaeli, Y. 1977, \apj, {\bf 218}, 323.

\bref
Rephaeli, Y. 1979, \apj, {\bf 227}, 364.

\bref
Rephaeli, Y., and Salpeter, E. E. 1980, \apj, {\bf 240}, 20.

\bref
Roettiger, K., Burns, J., and Loken, C. 1993, \apjlett, {\bf 407}, 53.

\bref
Rosat Scientific Data Center, 1989, {\it Rosat Mission Description},
NRA 89-OSSA-2, 10.3.

\bref
Sarazin, C. L. 1986, \rmp, {\bf 58}, 1.

\bref
Sarazin, C. L., and Bahcall, J. N. 1977, \apjsup, {\bf 34}, 451.

\bref
Takeda, H., Nulsen, P. E. J., and Fabian, A. C. 1984, \mnras, {\bf 208},
261.

\bref
Takhara, F., Ikeuchi, S., Shibazaki, N., and Hoshi, R. 1976, \progphys,
{\bf 56}, 1093.

\bref
Thomas, P. A., and Couchman, H. M. P. 1992, \mnras, {\bf 257}, 11.

\bref
Tsai, J. C., Katz, N., and Bertschinger, E. 1993, \apj, submitted.

\bref
Walker, T. P., Steigman, G., Schramm, D. N., Olive, K. A., and Kang, H.
1991, \apj, {\bf 376}, 51.

\bref
Watt, M. P., Ponman, T. J., Bertram, D., Eyles, C. J., Skinner, G. K.,
and Willmore, A.P. 1992, \mnras, {\bf 258}, 738.

\bref
White, R. E. 1991, \apj, {\bf 367}, 69.

\bref
White, S. D. M. 1976, \mnras, {\bf 177}, 717.

\bref
White, S. D. M. 1980, \mnras, {\bf 191}, 1P.

\bref
White, S. D. M., Navarro, J. F., Evrard, A. E., and Frenk, C. S. 1993,
\nature, submitted.

\bref
Yahil, A., and Ostriker, J. P. 1973, \apj, {\bf 185}, 787.

\pagebreak

\pagestyle{empty}
\centerline{
\begin{tabular}{lc}
\multicolumn{2}{c}{\bf Table 1} \\
\multicolumn{2}{c}{\bf Summary of Model Parameters} \\ \hline\hline
Parameter	& Value \\ \hline
Comoving Box Length ($\mpc$)			& 40 \\
Total Mass in Box ($\msol$)			& $4.44\times 10^{15}$\\
Number of Dark Matter Particles			& 32768	\\
Mass per Dark Matter Particle ($\msol$)		& $1.22\times 10^{11}$\\
Initial Number of Gas Particles			& 28733 \\
Mass per Gas Particle ($\msol$)			& $1.35\times 10^{10}$\\
Number of ``Galaxies''				& 108 \\
Comoving Gravitational Softening $\epsilon$ ($\kpc$) & 78 \\
Minimum Smoothing Length $h_{min}$ ($\kpc$)	& 110 \\
Initial Redshift				& 9 \\
Initial Mean Overdensity			& 0.031 \\
Timestep (yr)					& $1.33\times 10^{7}$\\
Initial Temperature of Gas			& $10^{4}\K$ \\
Temperature of Ejected Gas Before Averaging	& $1.37\times 10^{8}\K$ \\
\hline \\
\end{tabular}
}

\bigskip
\centerline{
\begin{tabular}{|l|c|c|c|c|c|}
\multicolumn{6}{c}{\bf Table 2} \\
\multicolumn{6}{c}{\bf Dark Matter Density Profile Fits} \\
\hline
Run	& Cutoff Radius	& $\rho_{0}$ & $r_{c}$ & $\alpha_{DM}$ &
 $\chi^{2} / $ \\
&($\mpc$) & ($\msol \mpc^{-3}$) & ($\kpc$) & & df \\
\hline\hline
& 1.0 & $\left(2.30\pm0.19\right)\times 10^{15}$ & $180\pm14$ &
 $0.823\pm0.032$ & 1.59 \\
EJ & 2.0 & $\left(2.53\pm0.20\right)\times 10^{15}$ & $151\pm9$ &
 $0.748\pm0.011$ & 2.07 \\
& 3.0 & $\left(2.24\pm0.16\right)\times 10^{15}$ & $179\pm8$ &
 $0.804\pm0.008$ & 3.06 \\
\hline
& 1.0 & $\left(4.57\pm0.49\right)\times 10^{15}$ & $100\pm9$ &
 $0.709\pm0.021$ & 1.51 \\
2F & 2.0 & $\left(4.47\pm0.43\right)\times 10^{15}$ & $102\pm7$ &
 $0.710\pm0.009$ & 2.44 \\
& 3.0 & $\left(3.48\pm0.30\right)\times 10^{15}$ & $135\pm7$ &
 $0.772\pm0.007$ & 5.42 \\
\hline
\end{tabular}
}

\bigskip
\centerline{
\begin{tabular}{|l|c|c|c|c|c|}
\multicolumn{6}{c}{\bf Table 3} \\
\multicolumn{6}{c}{\bf Gas Density Profile Fits} \\
\hline
Run	& Cutoff Radius	& $n_{0}$ & $r_{c}$ & $\alpha_{ICM} $ &
 $\chi^{2} / $ \\
&($\mpc$) & ($\cc$) & ($\kpc$) & & df \\
\hline\hline
& 1.0 & $\left(3.76\pm0.33\right)\times 10^{-3}$ & $312\pm44$ &
 $0.828\pm0.087$ & 0.76 \\
EJ & 2.0 & $\left(3.65\pm0.28\right)\times 10^{-3}$ & $311\pm21$ &
 $0.796\pm0.022$ & 1.38 \\
& 3.0 & $\left(3.26\pm0.22\right)\times 10^{-3}$ & $365\pm18$ &
 $0.868\pm0.015$ & 1.76 \\
\hline
& 1.0 & $\left(5.44\pm0.42\right)\times 10^{-3}$ & $331\pm37$ &
 $0.915\pm0.075$ & 0.35 \\
2F & 2.0 & $\left(5.63\pm0.38\right)\times 10^{-3}$ & $300\pm17$ &
 $0.842\pm0.019$ & 0.94 \\
& 3.0 & $\left(4.82\pm0.29\right)\times 10^{-3}$ & $372\pm16$ &
 $0.947\pm0.014$ & 1.99 \\
\hline
\end{tabular}
}

\bigskip
\centerline{
\begin{tabular}{|l|c|c|c|c|}
\multicolumn{5}{c}{\bf Table 4} \\
\multicolumn{5}{c}{\bf X--ray Surface Brightness Profile Fits} \\
\hline
Run & Redshift & $\Sigma_{0}$ ($\sbunits$) & $r_{x}$ ($\kpc$) & $\beta_{fit}$
\\ \hline\hline
& & & & \\
& $z\,=\,0.25$ & $\left(1.07\pm0.09\right)\times 10^{-2}$ &
 $839\pm301$ &
 $1.882\pm1.046$ \\
EJ & $z\,=\,0.10$ & $\left(1.80\pm0.10\right)\times 10^{-2}$ &
 $294\pm24$ & $0.681\pm0.042$ \\
& $z\,=\,0.06$ & $\left(3.01\pm0.07\right)\times 10^{-2}$ & $409\pm15$ &
 $0.929\pm0.032$ \\
& $z\,=\,0.02$ & $\left(3.88\pm0.04\right)\times 10^{-2}$ & $282\pm4$ &
 $0.741\pm0.008$ \\
\hline
& $z\,=\,0.25$ & $\left(2.54\pm0.14\right)\times 10^{-2}$ & $778\pm115$ &
 $1.776\pm0.376$ \\
2F & $z\,=\,0.10$ & $\left(5.68\pm0.17\right)\times 10^{-2}$ & $295\pm12$ &
 $0.769\pm0.024$ \\
& $z\,=\,0.06$ & $\left(8.99\pm0.15\right)\times 10^{-2}$ & $267\pm5$ &
 $0.734\pm0.010$ \\
& $z\,=\,0.02$ & $\left(8.99\pm0.07\right)\times 10^{-2}$ & $271\pm2$ &
 $0.788\pm0.005$ \\
\hline
\end{tabular}
}

\bigskip
\centerline{
\begin{tabular}{|c|c|c|c|}
\multicolumn{4}{c}{\bf Table 5} \\
\multicolumn{4}{c}{\bf $\beta_{spec}$ Values} \\
\hline
Redshift	&	Parameter	& EJ	& 2F \\ \hline\hline
& & & \\
$z\,=\,0.25$ & $\beta_{DM}$ & 1.09 & $1.20$ \\
 & $\beta_{GAL}$ & 0.96 & --- \\
\hline
$z\,=\,0.10$ & $\beta_{DM}$ & 1.06 & 1.25 \\
 & $\beta_{GAL}$ & $0.79$ & --- \\
\hline
$z\,=\,0.06$ & $\beta_{DM}$ & 1.05 & 1.06 \\
 & $\beta_{GAL}$ & 0.79 & --- \\
\hline
$z\,=\,0.02$ & $\beta_{DM}$ & 0.92 & 1.14 \\
 & $\beta_{GAL}$ & 0.68 & --- \\
\hline
\end{tabular}
}

\pagebreak
\pagestyle{empty}
\centerline {Appendix}
\bigskip

The emissivity due to free-free encounters between electrons and ions in a
fully ionized primordial plasma ($X \se 0.76$, $Y \se 0.24$) with temperature
$T$ and particle density $n$ is given by
(Spitzer 1968)
$$
\epsilon_{\nu} \ = \ 4.32 \times 10^{-28} \ n^2 \ T^{1/2} \ \ (h/kT) \
\bar g(h\nu/kT) \ e^{-h\nu/kT} \ergs \cc \, {\rm Hz}^{-1}
\eqno{(A1)}
$$
where $\bar g(h\nu/kT)$ is an averaged Gaunt factor providing quantum
mechanical corrections to the classical expression.  Define a bolometric
emissivity using unit Gaunt factor
$$
\epsilon_{bol} \ = \ \int_0^\infty \, d\nu \ \epsilon_{\nu}
 \ = \ 4.32 \times 10^{-28} \ n^2 \ T^{1/2} \ \ergs \cc .
\eqno{(A2)}
$$
Then the energy radiated within a given energy band $E_1-E_2$ can be
expressed as
$$
\epsilon_{band} \ = \ f_{band}(T) \ \epsilon_{bol}
\eqno{(A3)}
$$
where
$$
f_{band}(T) \ = \ \int_{E_1/kT}^{E_2/kT} \, d\eta \ \bar g(\eta) \,
e^{-\eta}  .
\eqno{(A4)}
$$
The band limited x-ray emission from a given volume $V$ is found by
$$
L_x \   = \ \int_V \, d^3x \, \epsilon_{band}(x)
        = \ 4.32 \times 10^{-28}
        \int_V \, d^3x \  n^2(x) \ T^{1/2}(x) \ f_{band}(T(x)) \ \ergs .
\eqno{(A5)}
$$
Using the discrete information in the simulations, $\int_V \, d^3x
\rho \rightarrow \Sigma_i m_i$, implies
the x-ray luminosity from a given  volume can be
calculated by
$$
L_x \ = \ 4.32 \times 10^{-28} \ (\mu m_p)^{-2} \ m_g \
        \sum_i \rho_i \, T_i^{1/2} \, f_{band}(T_i) \ergs
\eqno{(A6)}
$$
where the sum runs over particles within the volume.
The flux $F_x$ of x-rays at earth in an observed energy band $E_1 - E_2$ is
(Weinberg 1972)
$$
F_x \ = \ (4\pi)^{-1} \ \biggl( {\Ho \over 2c} \biggr)^2 \
\biggl( {L_{bol} \,  f_{band}(T,z) \over (1+z)^2(1-(1+z)^{-1/2})^2 } \biggr)
\eqno{(A7)}
$$
where $f_{band}(T,z)$ is the integral given by equation (A4) over an
appropriately redshifted energy range $E_1(1+z)$ to $E_2(1+z)$.

\pagebreak
\section{Figure Captions}

\ii{Figure 1} Evolution of the fluids in the simulation.  Each frame
shows the particles contained within a slice of comoving dimensions
33.5 Mpc by 33.5 Mpc wide, 6.7 Mpc thick,
projected along the z--axis.  Each row corresponds to a given redshift,
which is shown to the right of the panels.  The first two columns show
the dark matter and gas distributions, sampling $1/7$ of the particles
for clarity.  The third column shows the distribution of
all ejected gas particles in the slice; it is thus a subset of the
full (not sparsely sampled) gas distribution.  The fourth column shows the
location of the galaxy particles within the slice.

\ii{Figure 2} Temperature profile of the cluster gas for the ejection
(EJ) and two--fluid (2F) runs.  The profiles are centered on the potential
minimum of the cluster, with the cluster at any epoch defined as the
most massive progenitor of the final cluster.  The five lines
correspond to different redshifts:  $z\,=\,0.02$,
solid; $z\,=\,0.102$, dotted;
$z\,=\,0.246$, short dashed; $z\,=\,0.499$, long dashed; and
$z\,=\,1.01$, dash--dot.

\ii{Figure 3} Density profiles of the dark matter, gas and galaxies.
Data bins are Lagrangian; they are defined by 200 particles per
bin for the dark matter and gas, and by 5 particles per bin for
the galaxies.

\ii a) and b) Least squares best--fits of the true dark matter
density profile of the cluster, at $z\,=\,0.02$, to
Equation~\ref{eq:betaden}.  Left plot is from the ejection run;
right plot is from the two--fluid run.  The smallest radial point
was not used in calculating the fit.  Three
lines are present in each plot, corresponding to fits where data points
beyond $1\mpc$ (solid), $2\mpc$ (dotted), and $3\mpc$ (dashed) were
excluded from the fit.  The primary difference between the three lies
in the large radius slope (value of $\alpha$).

\ii c) and d) Least squares best--fits of the true gas number density
profile of the cluster at $z\,=\,0.02$, to Equation~\ref{eq:betaden}.
Data plotted as for the dark matter, in a) and b).

\ii e) and f) Galaxy number density profile of the cluster
at $z\,=\,0.02$ (e) and at $z\,=\,9.0$ (f).  Least squares best--fits
of the $z\,=\,0.02$ data to Equation~\ref{eq:betaden} are shown, using
the same fitting constraints as in a)--d) above.  The ``center'' of the
cluster at high redshift is determined by identifying the dark matter
making up the cluster at $z\,=\,0.02$, and using the center of that
set of particles' distribution at $z\,=\,9.0$.

\ii{Figure 4} Entropy profile of the cluster gas for the two
runs, evaluated at the same redshifts as Figure 2 with the same
linestyles.  The appearance of a jump in central entropy for the
$z\,=\,0.246$ profile is a consequence of centering on the most
bound dark matter particle; at that epoch, the potential minimum is
briefly offset from the low entropy gas because of merger
activity, as described in the text.

\ii{Figure 5} The iron abundance profile of the cluster, at four
different redshifts:  $z\,=\,0.02$ (solid), $z\,=\,0.102$ (dotted),
$z\,=\,0.499$ (short dashed),  and $z\,=\,1.01$ (long dashed).

\ii{Figure 6} Value of three mass ratios plotted against enclosed
mass (against radius which encloses that mass):  mass in galaxies
over mass in (necessarily primordial) gas at $z\,=\,9.0$ (solid);
mass in galaxies over mass in primordial gas at $z\,=\,0.02$
(dotted); mass in ejected gas over mass in primordial gas at
$z\,=\,0.02$ (dashed).

\ii{Figure 7} Evolutionary sequence of iron abundance profiles
for three proto--cluster objects which merge over the time
period shown.  The solid line corresponds to the largest progenitor
of the three at $z\,=\,0.465$, the dashed line to the smallest.
Lines disappear when they merge into the more massive object.
Merging does not strongly affect the abundance gradient.

\ii{Figure 8} Four plots describing the projected, flux--weighted
iron abundance of the simulated cluster.  The upper left hand corner
shows a $64^\prime$ square map of the apparent iron
abundance created by assuming
ROSAT PSPC resolution ($0^\prime .5$) at all locations in
the map.  The contour levels used are at abundances of 0.4, 0.8,
1.2, 1.6, and 2.0 times solar.  The upper right hand corner
shows discrete sampling of this map
offset from the x--ray surface brightness maximum by the angle
and in the direction shown.  The lower left hand corner shows
an apparent iron abundance map generated by assuming EXOSAT
observations at all locations in the map, with an instrument
response modelled by a Gaussian with a FWHM of 45'; to the right
are the profiles that such observations would generate.  The
contour levels used in the map are at 0.8, 0.85, 0.9, 0.95, 1.0,
1.05, and 1.07 times solar.  The dots in the maps correspond to
the locations of galaxies in this projected field of view; the
X shows the location of the X--ray surface brightness maximum.

\ii{Figure 9} Evolution of $L_{x}$, within a comoving 1 Mpc,
band--limited to
the ROSAT PSPC passband, with $z$, for ejection run (solid line)
and two--fluid run (dotted line).

\ii{Figure 10} Evolutionary sequence of simulated ROSAT PSPC x--ray
images for both runs, in z--axis projection at the indicated
redshifts.  a) is ejection run; b) is two--fluid run.
Each image is in a $64^\prime$ field of view, with $0^\prime.5$
resolution.  The contour levels used (in units of
$10^{-4} \sbunits$) are at 3.16, 6.31, 12.6, 25.1, 50.1, 100, 200,
398, and 794 (logarithmic separations of 0.3).  The dots in a) correspond
to the projected positions of galaxies within the field of view.
Because of its lower central gas entropy, the two--fluid run is
visible to higher redshift that the ejected run.
The source in the northwest of the images at $z\,=\,0.499$ and
$z\,=\,0.246$ in the 2F is still separate (has not merged) at the
redshifts of the later images; it is outside the field of view at those
redshifts.

\ii{Figure 11} Surface brightness profiles for both runs, at the four
lowest redshifts imaged in Figure 10.  Also shown are fits to the
$\beta$-model.  No data is shown where the surface brightness drops
below the fiducial choice for the background of $3\times10^{-4}\sbunits$.
Table 4 gives the best--fit parameter values and errors.

\ii{Figure 12} Binding and gas mass within radius (crosses and open squares),
along with projections of hydrostatic isothermal $\beta$--model based on
x--ray data, for the ejection run at a redshift $z\,=\,0.02$.  X--ray
surface brightness fitting parameters used are from the fit to the z--axis
projection at that redshift.  Temperature is the isothermal central
temperature for that redshift from Figure 2.  Corrections for the
softened force law used in the simulations are included.

\ii{Figure 13} The baryon fraction of all matter within radius $r$, for
the ejection and two--fluid runs.  The overall value of the baryon
fraction for the simulations, 0.1, is shown as a dotted line.  The
dashed lines indicate the true baryon fraction profiles, while the
solid lines represent predictions for each simulation based on
mass estimates from the hydrostatic isothermal beta model.

\ii{Figure 14} Redshift evolution of the true values of $\beta$,
calculated using the mean temperature and 1--D velocity dispersion within
a comoving radius of 1 Mpc.  The solid line shows $\beta_{DM}$
using the dark matter's velocity dispersion; the dotted line shows
$\beta_{gal}$ with a 1--D velocity dispersion calculated from the
galaxies within 1 Mpc.  The dashed line shows the velocity bias
factor, expressed as the ratio of the two velocity dispersions,
$b_{v} \,=\, \sigma_{gal}/\sigma_{DM}$.  Some values from this plot
are shown in Table 5.  A persistent velocity bias is evident which
leads to $\beta_{gal} < 1$.

\pagebreak

\end{document}